\documentclass[journal,12pt,onecolumn,draftclsnofoot]{IEEEtran}

\ifCLASSINFOpdf

\else

\fi
\usepackage[colorinlistoftodos]{todonotes}
\usepackage{amssymb}
\usepackage{amsmath,epsfig,amssymb,verbatim}
\usepackage{cite}
\usepackage[caption=false]{subfig}
\usepackage{array,algorithm,algorithmic}
\usepackage{comment}
\def\BibTeX{{\rm B\kern-.05em{\sc i\kern-.025em b}\kern-.08em
    T\kern-.1667em\lower.7ex\hbox{E}\kern-.125emX}}

\usepackage{enumitem}
\setlist{leftmargin=5.5mm}
\usepackage{enumitem}
\setlist{leftmargin=4mm}
\usepackage{tablefootnote}

\newtheorem{lemma}{Lemma}

\begin{document}
\vspace{-5mm}
\title{Spectrum Allocation with Adaptive Sub-band Bandwidth for Terahertz Communication Systems}

\author{Akram~Shafie,~Nan~Yang,~Sheeraz~Alvi,~Chong~Han, Salman~Durrani,~and~ Josep~M.~Jornet
\thanks{Akram Shafie, Nan Yang, Sheeraz Alvi, and Salman Durrani are with the School of Engineering, The Australian National University, Canberra, ACT 2601, Australia (e-mail: akram.shafie@anu.edu.au, nan.yang@anu.edu.au, sheeraz.alvi@anu.edu.au, salman.durrani@anu.edu.au).}
\thanks{Chong Han is with the UM-SJTU Joint Institute, Shanghai Jiao Tong University, Shanghai 200240, China (e-mail: chong.han@sjtu.edu.cn). }
\thanks{Josep~M.~Jornet is with the Department of Electrical and Computer Engineering, Northeastern University, Boston, MA 02120, USA (e-mail: jmjornet@northeastern.edu). }
}

\markboth{This work has been accepted for publication in IEEE Transaction on Communications.}{Yuan \MakeLowercase{\textit{et al.}}: Coverage Analysis for 3D Terahertz Communication Systems}

\maketitle
\vspace{-10mm}

\begin{abstract}
We study spectrum allocation for terahertz (THz) band communication (THzCom) systems, while considering the frequency and distance-dependent nature of THz channels. Different from existing studies, we explore multi-band-based spectrum allocation with adaptive sub-band bandwidth (ASB) by allowing the spectrum of interest to be divided into sub-bands with unequal bandwidths. Also, we investigate the impact of sub-band assignment on multi-connectivity (MC) enabled THzCom systems, where users associate and communicate with multiple access points simultaneously. We formulate resource allocation problems, with the primary focus on spectrum allocation, to determine sub-band assignment, sub-band bandwidth, and optimal transmit power. Thereafter, we propose reasonable approximations and transformations, and develop iterative algorithms based on the successive convex approximation technique to analytically solve the formulated problems. Aided by numerical results, we show that by enabling and optimizing ASB, significantly higher throughput can be achieved as compared to adopting equal sub-band bandwidth, and this throughput gain is most profound when the power budget constraint is more stringent. We also show that our sub-band assignment strategy in MC-enabled THzCom systems outperforms the state-of-the-art sub-band assignment strategies and the performance gain is most profound when the spectrum with the lowest average molecular absorption coefficient is selected during spectrum allocation.
\end{abstract}

\vspace{-3mm}
\begin{IEEEkeywords}
Terahertz communication, spectrum allocation, multi-connectivity, adaptive bandwidth.
\end{IEEEkeywords}
\section{Introduction}

The scarcity of spectral resources in contemporary wireless communication systems to meet the  unprecedented increase in wireless data traffic in the next decade, has advocated the investigation of suitable regimes in the electromagnetic spectrum  for the sixth-generation (6G) and beyond~\cite{2020_Mag6G_Marco_UseCasesandTechnologies}.
Against this backdrop, the ultra-wide terahertz (THz) band ranging from 0.1 to 10 THz has recently attracted considerable attention from the wireless communication research community~\cite{2018MagCombatDist}. The huge available bandwidths in the order of tens up to a hundred gigahertz (GHz) offer enormous potential to support emerging wireless applications that demand an explosive amount of data, such as holographic telepresence, augmented reality, virtual reality, and wireless backhaul.
Built on the major progress in  THz band channel modeling and standardization efforts over the past decade, it is anticipated that indoor THz band communication (THzCom) systems will be brought to reality in the near future.

The exploration of novel and efficient spectrum allocation strategies is of paramount significance to harness the potentials of the THz band~\cite{2020_WCM_THzMag_TerahertzNetworks,2020_WCM_THzMag_Standardization}. When such strategies are to be devised, the unique characteristics of the THz band  pose new and pressing challenges that have never been seen at lower frequencies~\cite{2011_Jornet_TWC}. Specifically, in addition to the severe spreading loss and the higher channel sparsity, the THz band is characterized by the unique molecular absorption loss which is frequency and distance-dependent. On one hand, the molecular absorption loss divides the entire THz band into multiple ultra-wide THz transmission windows (TWs). On the other hand, it introduces drastically varying path loss even within a specific THz TW and this variation increases further as the transmission distance increases.
In addition to these, limited advancements in THz band digital processors, frequency synchronizers, and transceivers make it very challenging to apply the complex spectrum reuse and multiplexing strategies that are suggested for sub-6 GHz and mmWave systems into the THz band~\cite{Chong2019Archive}.
Thus, the design of novel, low-complexity, and efficient spectrum allocation strategies is of utmost importance to  develop ready-to-use THzCom systems.

\subsection{Related Studies and Motivation}

\textcolor{black}{In the literature, two types of  carrier-based spectrum allocation schmes have been studied for micro- and macro-scale multiuser THzCom systems, namely, multi-TW-based spectrum allocation and multi-band-based spectrum allocation. In the multi-TW-based spectrum allocation scheme, individual TWs are fully allocated to separate high-speed communications links while exploring wideband signals that have bandwidths equal to those of TWs~\cite{akramICC2020,HBM1}.
When this scheme is adopted in multiuser THzCom systems, the same spectrum needs to be shared among multiple users, since the number of available TWs within the entire THz band is limited~\cite{HBM1}. This necessitates the exploration of spatial and temporal multiplexing strategies~\cite{HBM1}. In addition, broadening and channel squint effect mitigation techniques, as well as efficient beamforming and medium access control (MAC) protocols, need to be designed to overcome interference.
In the multi-band-based spectrum allocation scheme, the spectrum of interest is divided into a set of non-overlapping sub-bands that have a relatively small bandwidth, and then the sub-bands are utilized to satisfy the service demands of single or multiple users in the system~\cite{HBM3,HBM2}.}
\textcolor{black}{When this scheme is adopted in multiuser THzCom systems, the users can be assigned to different sub-bands to ensure intra-band interference-free data transmission, as there is a relatively larger number of sub-bands. It is also noted that this scheme efficiently allocates spectral resources when there is high absorption loss variation among the links in multiuser systems.} Thus, this scheme has been widely explored for micro- and macro-scale multiuser THzCom applications. In this work, we focus on multi-band-based spectrum allocation for multiuser THzCom systems.

The first study on multi-band-based spectrum allocation was presented in~\cite{HBM3} and  the impacts of distance-varying usable bandwidth and different types of interference on multi-band-based spectrum allocation were discussed.
Moreover, a distance-aware multi-carrier (DAMC) based sub-band assignment strategy for multi-band-based spectrum allocation was proposed  in~\cite{HBM2} to improve the throughput fairness among users in  multiuser THzCom systems. In~\cite{HBM2}, it was proposed to assign the sub-bands that exist at the edges of the THz TW to the links with longer distances and the sub-bands that exist in the center region of the THz TW to the links with shorter distances, in order to take advantage of the frequency and distance-dependent nature of THz channels.
\textcolor{black}{Adopting DAMC-based spectrum allocation, resource allocation problems in non-orthogonal multiple access (NOMA) assisted THzCom system and THz band backhaul network were discussed in~\cite{2019_Chong_DABM2} and~\cite{2020_Chong_InfoCom_DABM}, respectively.}
\textcolor{black}{In other studies,  efficient sub-band assignment and transmit power algorithms based on the  alternative direction method and $K$-means clustering were developed  in \cite{2021_THz_NOMA} and \cite{2020ICC_NOMAforTHz}, respectively, for  NOMA-assisted THzCom systems.}

Note that the previous studies in~\cite{HBM3,HBM2,2020_Chong_InfoCom_DABM,2019_Chong_DABM2,2021_THz_NOMA,2020ICC_NOMAforTHz} that analyzed multi-band-based spectrum allocation all considered equal sub-band bandwidth (ESB), where the spectrum of interest is divided into sub-bands with equal bandwidth. However,   
 it would be beneficial to explore spectrum allocation with adaptive sub-band bandwidth (ASB) to improve the spectral efficiency, by allowing the spectrum of interest to be divided into sub-bands with unequal bandwidths. Specifically, given that the absorption loss varies considerably within THz TWs, the absorption loss variation among the sub-bands would be very high when ESB is considered~\cite{HBM3,HBM2}. The impact of this high variation may possibly be mitigated by adaptively adjusting the sub-band bandwidths, which leads to an overall improvement in the throughput performance. This is one of the motivations of this work.
\textcolor{black}{We clarify that the studies in~\cite{akramICC2020,HBM1} considered variable bandwidths for THzCom systems. However, we note that~\cite{akramICC2020,HBM1} explored the multi-TW-based spectrum allocation scheme. Considering variable bandwidth in this scheme, the bandwidths  of communication links vary according to the variation of the usable bandwidth of THz TWs. This is different from the ASB principle in our multi-band-based spectrum allocation.}

\textcolor{black}{A key factor to consider during THz band spectrum allocation is the impact of several 6G-enabling technologies, such as multi-connectivity (MC) and intelligent reflective surfaces (IRSs), which  have  been envisioned to be integrated into THzCom systems~\cite{2020_Chong_IRSTHz2,2020_WCM_THzMag_TerahertzNetworks,2021_THzMCHandOver}. In this work, we focus on an MC-enabled THzCom system.}
MC allows users to associate and communicate with multiple access points (APs) simultaneously\footnote{We clarify that several terminologies such as multi-connectivity, cooperative transmission,  joint transmission, concurrent transmission, and coexisting communication, have been used in the literature to denote simultaneous association and communication  of users with multiple APs~\cite{2020_TC_MLApproach_GeoffreyYeLi,2020_IEEESurveyandTutorials_MCforURLLC,2018_TC_Idea}. However, following~\cite{2020_TC_MLApproach_GeoffreyYeLi,2020_IEEESurveyandTutorials_MCforURLLC}, we use the term multi-connectivity in this work.}~\cite{2020_TC_MLApproach_GeoffreyYeLi,2020_IEEESurveyandTutorials_MCforURLLC,2018_TC_Idea},  thereby enabling to overcome the performance degradation caused by the high vulnerability of THz signals towards blockages.
With careful design, intra-band  or inter-band MC strategies can be utilized to enhance the reliability and/or throughput of THzCom systems, depending on the application scenario. Here, intra-band MC refers to the utilization of one or more spectra in the THz band for association  and communications~\cite{akramICC2020,2017N5}, while inter-band MC (or multiple radio access technology) refers to the utilization of spectra in both the THz band and the sub-6 GHz and/or mmWave~\cite{2016_Jornet_MultiRAT,2020_WCL_Multi_RATMCforTHz}.
In this work, we focus on  intra-band MC due to its ability to support high data rate transmission even when the primary associated link is blocked by blockers.

\textcolor{black}{Spectrum allocation in an intra-band MC-enabled THzCom system was  discussed ~\cite{akramICC2020}. However, we note that~\cite{akramICC2020} considered multi-TW-based spectrum allocation for a single user and its primary focus was on analyzing the impact of different types of MC strategies.} Moreover, the resource allocation in an intra-band MC-enabled multiuser THzCom system was discussed in~\cite{2017N5}. However,  the impacts of blockages and  spectrum allocation with ASB were not considered in~\cite{2017N5}. Furthermore, it is noted that the previous studies on multi-band-based spectrum allocation in~\cite{HBM3,HBM2,2020_Chong_InfoCom_DABM,2019_Chong_DABM2,2021_THz_NOMA,2020ICC_NOMAforTHz} did not consider the impacts of blockages, nor MC strategies. Thus, there exists the need to investigate multi-band-based spectrum allocation with ASB for MC-enabled multiuser THzCom systems.  

\enlargethispage{3mm}
\subsection{Our Contributions}

In this work, we study spectrum allocation for intra-band MC-enabled multiuser THzCom systems,  when users associate and communicate with multiple access points simultaneously.
Specifically, we focus on  multi-band-based spectrum allocation under the consideration that  the associated links in the system are served by separate sub-bands. 
 The main contributions of this work are summarized as follows:

\begin{itemize}
  \item We propose multi-band-based spectrum allocation with ASB to improve the spectral efficiency and study the impact of sub-band assignment  on MC-enabled multiuser THzCom systems. To this end, we formulate a  generalized resource allocation problem with the objective of maximizing the throughput of the MC-enabled multiuser THzCom system, while primarily focusing on spectrum allocation. This problem consists of sub-band assignment, user association identification, sub-band bandwidth allocation, and power control. 
      Although the generalized resource allocation problem can significantly improve the performance of the considered THzCom system when the spectrum of interest exists anywhere within a THz TW, we find that  it is extremely difficult, if not impossible, to analytically solve it by using traditional optimization theory techniques. Due to this, we simplify some of the constraints in the generalized resource allocation problem and obtain two modified problems that also correspond to realistic scenarios of the considered THzCom system.
  \item We formulate the  first modified problem, which is \emph{the resource allocation with ESB}, while considering that the spectrum of interest is divided into sub-bands with equal bandwidths. The novelty of the resource allocation with ESB lies in the consideration of sub-band assignment in MC-enabled multiuser THzCom systems, while the previous relevant studies on THz band spectrum allocation have considered DAMC-based sub-band assignment, the optimality of which for MC-enabled multiuser THzCom systems needs to be validated.  We also formulate the second modified problem, which is \emph{the resource allocation with ASB in either a positive absorption coefficient slope region (PACSR) or a negative absorption coefficient slope region (NACSR)}, while considering that the spectrum of interest fully exists in either a PACSR or an NACSR of the THz band. We define PACSRs and NACSRs as the regions with the increasing and decreasing absorption coefficient, respectively, within the THz TWs. The novelty of the resource allocation with ASB in a PACSR/NACSR lies in the consideration of \textcolor{black}{multi-band-based} spectrum allocation with ASB, while, to the best of our knowledge, recent studies on \textcolor{black}{multi-band-based} spectrum allocation for THzCom systems have only considered ESB.
  \item  To analytically solve the resource allocation with ESB, which is a mixed-integer nonlinear problem, we introduce transformations and propose an iterative algorithm based on the successive convex approximation (SCA) technique. Also, we propose reasonable approximations and transformations to the resource allocation with ASB in a PACSR/NACSR to arrive at an approximate convex problem. Thereafter, we develop an iterative algorithm based on the SCA technique to solve the approximate convex problem.
  \item Aided by numerical results, we show that the proposed resource allocation with ESB outperforms the DAMC-based  spectrum allocation in MC-enabled multiuser THzCom systems. The performance gain is most profound when the spectrum with the lowest average molecular absorption loss within the THz TW is selected during spectrum allocation and when the number of APs with which each user associates is high. Our results also show that the  proposed resource allocation with ASB in a PACSR/NACSR achieves a significantly higher throughput performance  (between $13~\%$ and $33~\%$) compared to the spectrum allocation strategies that consider ESB, due to its enhanced ASB capability. The performance gain is most profound when the  power budget constraint is more stringent and when the upper bound on sub-band bandwidth is very high.
\end{itemize}

\enlargethispage{3mm}
The rest of the paper is organized as follows. In Section II, we describe the system model. In Section III, we formulate the generalized resource allocation problem and introduce the two modified problems. In Section IV, we formulate and solve the resource allocation problem with ESB. In Section V, we formulate and solve the resource allocation problem with ASB in a PACSR/NACSR. Numerical results are presented in Section VI. Finally, we conclude the paper in Section VII.

\section{System Model}

In this section, we first introduce the system model considered in this work, by describing the system deployment, the THz spectrum, and the channel model. Thereafter, we specify the performance metrics of interest.

\subsection{System Deployment}

\begin{figure*}[t]
\centering
\includegraphics[width=0.99\columnwidth]{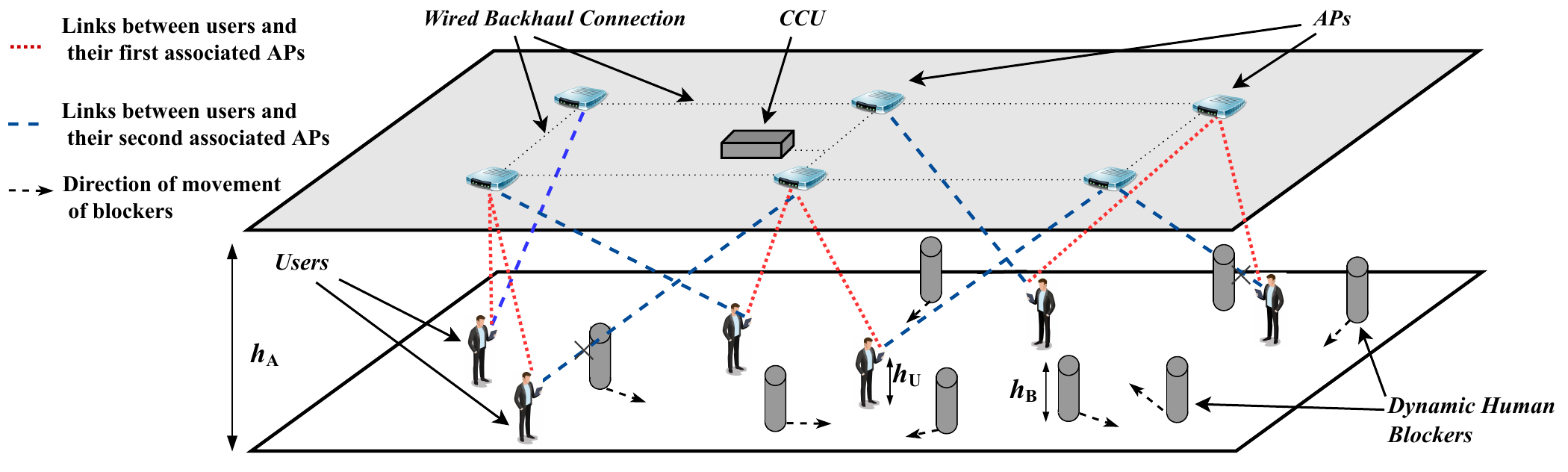}
\vspace{-7mm}
\caption{Illustration of the considered 3D MC-enabled multiuser THzCom system where 6 stationary users associate with 6 fixed APs  with the MC order of 2, in the presence of dynamic blockers.}\label{Fig:SystemModel}
\vspace{-3mm}
\end{figure*}

In this work, we consider a cell free architecture based  three-dimensional (3D) indoor THzCom system, as depicted in Fig.~\ref{Fig:SystemModel}, where $J$ APs cooperatively support the uplink of $I$ stationary users which demand high data rates. We consider that APs are mounted on the ceiling of the indoor environment; thus are modeled as having the same fixed height of $h_{\textrm{A}}$.
We denote $\mathcal{J}= \{1, 2, \cdots , j, \cdots, J\}$ as the set of these APs
and assume that their location follows a particular layout, e.g., the layout  corresponding to the indoor environment specified in the 3GPP standard~\cite{3GPPStand}. We also consider that the users are distributed uniformly in the indoor environment. We denote $\mathcal{I}= \{1, 2, \cdots , i, \cdots , I\}$ as the set of the users and assume them to be of fixed height $h_{\textrm{U}}$.
We denote $r_{ij}$ and $d_{ij}=\sqrt{(h_{\textrm{A}}{-}h_{\textrm{U}})^2 {+}r_{ij}^2}$ as the horizontal and 3D distances of the link between the $i$th user and the $j$th AP, respectively~\cite{akram2020JSAC}.


The extremely short wavelengths of THz signals make the THz signal propagation
to be highly vulnerable to blockages~\cite{2011_Jornet_TWC,akram2020JSAC}.
In our system, we consider moving humans act as potential impenetrable blockers. These blockers are modeled as cylinders with the radius of $r_{\textrm{B}}$ and height of $h_{\textrm{B}}$ and their location follows a Poisson point process (PPP) with the density of $\lambda_{\textrm{B}}$~\cite{akramICCWS2020}. We assume that the mobility of blockers follows the random directional model  with constant moving speed of $v_{\textrm{B}}$~\cite{RDM1,MC4}.
For the sake of practicality, we assume that $h_{\textrm{A}} > h_{\textrm{B}} > h_{\textrm{U}}$ in the considered system.

We assume that users are enabled with the MC strategy of order $N$ to overcome the performance degradation caused by blockages~\cite{MC4,akramICC2020,2020_TC_MLApproach_GeoffreyYeLi,2020_IEEESurveyandTutorials_MCforURLLC,2018_TC_Idea}. Under the MC strategy of order $N$, each user associates and
communicates with $N$ APs, out of the possible $J$  APs in the system, simultaneously for user session continuity ($N\leqslant J$)~\cite{2020_TC_MLApproach_GeoffreyYeLi,2020_IEEESurveyandTutorials_MCforURLLC,2018_TC_Idea}. We also assume that all APs are connected to a central control unit (CCU) via wired backhaul connection.

\subsection{THz Spectrum}

\begin{figure*}[!t]
\centering
\includegraphics[width=1\columnwidth]{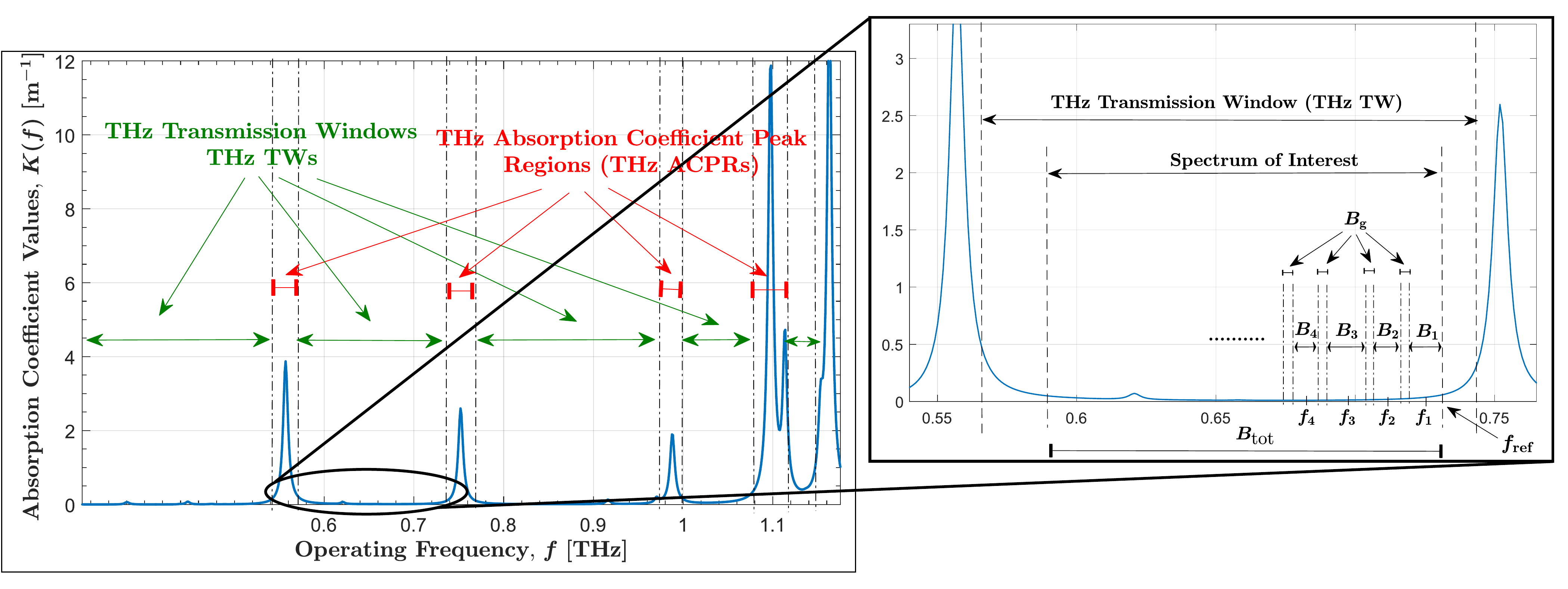}
\vspace{-5mm}
\caption{\textcolor{black}{Illustration of THz transmission windows, THz absorption coefficient peak regions, and the arrangement of sub-bands.}}\label{Fig:THzBand} \vspace{-3mm}
\end{figure*}

The intermittent absorption loss peaks, that are observed throughout the THz band at different frequencies, divide the entire THz band into THz absorption coefficient peak regions (ACPRs) and  ultra-wideband  THz transmission windows (TWs), as shown in Fig. \ref{Fig:THzBand}~\cite{2011_Jornet_TWC}. 
It is envisioned that THz TWs, rather than THz ACPRs, can be utilized for applications that demand ultra-high data rates, since molecular absorption loss is relatively low in THz TWs, but very high in THz ACPRs~\cite{2021_JSAC_VariableBandwidth}.
Considering this, we focus on the allocation of the spectrum that exist within THz TWs in this work.

As mentioned in Section I, in this work we focus on multi-band-based spectrum allocation with ASB. Thus, we consider that the spectrum of interest is divided into $S$ sub-bands  with unequal bandwidths, as shown in Fig. \ref{Fig:THzBand}. Also, we consider that these sub-bands are separated by guard bands that are of fixed  bandwidth~\cite{HBM2,HBM3,2011_TWC_IBISuppression}.
We denote $\mathcal{S} =
\{1, 2, \cdots, s, \cdots , S\}$ as the set of the sub-bands
and further denote $\mathcal{F}_{\mathcal{S}} =
\{f_1, f_2, \cdots  , f_s, \cdots , f_S\}$ and $\mathcal{B}_{\mathcal{S}}=\{B_1, B_2,$ $
 \cdots , B_s,\cdots , B_S\}$ as the sets of their center frequency and bandwidth, respectively. We note that 
\begin{align}\label{Equ:BmaxConst}
0\leqslant B_{s} \leqslant B_{\textrm{max}}, ~~~~~~ \forall s\in \mathcal {S}, \end{align}
 where $B_{\textrm{max}}$  denotes the upper bound on the sub-band bandwidths. We then denote $B_{\textrm{g}}$ as the fixed bandwidth of guard bands and $B_{\textrm{tot}}$ as the total available bandwidth within the spectrum of interest. Considering this,  we obtain
\begin{align}\label{Equ:BtotConst} \sum_{s\in \mathcal {S}}B_{s}+(S-1)B_{\textrm{g}}= B_{\textrm{tot}}. \end{align}
For notational convenience, we consider that sub-bands are labeled such that $f_1> f_2> \cdots > f_S$.
Thus, we have 
\begin{align}\label{Equ:f_s}
  f_{s}&=f_{\textrm{ref}} - \sum_{k=1}^{s-1}(B_{k}+B_{\textrm{g}})-B_{s}/2, ~~~~~~ \forall s\in \mathcal {S},
\end{align}
where $f_{\textrm{ref}}$  is the end-frequency of the spectrum of interest as shown in Fig. \ref{Fig:THzBand}.

\subsection{Channel Model}
\label{Sec:Channel}

The signal propagation at the THz band is determined by spreading and molecular absorption losses. Considering this, the channel transfer function is obtained as~\cite{2011_Jornet_TWC}
\begin{align} \label{Equ:ChannelOrg}
H(f,d) &= \frac{c}{4 \pi f d} ~\! e^{-\frac{1}{2}K(f) d},
\end{align}
where $c$ is the speed of light, $f$ is the frequency, $d$ is the distance of interest, and $K(f)$ is the molecular absorption coefficient at $f$.
In \eqref{Equ:ChannelOrg}, $e^{-\frac{1}{2}K(f) d}$ represents the molecular absorption loss  which is the result of oxygen molecules and water vapor absorbing THz signal energies for their rotational transition energies.
We clarify that $K(f)$ is calculated using the values from the HITRAN database for the given pressure, temperature, and humidity setting\footnote{The values of $K(f)$ are obtained from
$K(f)=\frac{p}{p_{\textrm{STP}}}\frac{T_{\textrm{STP}}}{T}\sum_{i,g}Q^{i,g}\sigma^{i,g}(f)$, where $p_{\textrm{STP}}$ and $T_{\textrm{STP}}$ are the standard pressure and temperature, respectively, and $p$ and $T$ are the pressure and the temperature of the transmission environment, respectively~\cite{2011_Jornet_TWC}. Also, $Q^{i,g}$ and $\sigma^{i,g}(f)$ are the total number of molecules per unit volume and the absorption cross section for the isotopologue $i$ of gas $g$ at the frequency $f$, respectively, and they are obtained from the HITRAN database~\cite{2008_Hitran}.}~\cite{2008_Hitran}.

At the THz band, non-line-of-sight (NLoS) rays are substantially attenuated due to high reflection and scattering losses, as well as blockages.
Due to these, NLoS rays are typically $15$-$20~\textrm{dB}$ weaker than that of line-of-sight (LoS) rays~\cite{2011_Jornet_TWC,2017_ChongTVT_Graphene,akram2020JSAC}. Hence, we ignore the impact of NLoS rays and focus only on the LoS rays of THz signals.

\subsection{Performance Metrics of Interest}

\subsubsection{Non-blockage Probability}

If a blocker is moving according to the random directional model in a given area, the probability density function of the location of blockers is uniform over time. As such, the location of blockers forms PPP with the same intensity of $\lambda_{\textrm{B}}$ at any given time instant~\cite{RDM1,akramICCWS2020}. Considering this,  the non-blockage probability of the link between the $i$th user and the $j$th AP can be derived as
\begin{equation}\label{Equ:pLH}
p_{\textrm{nb}}(r_{i,j})=\zeta e^{-\eta_{\textrm{B}} r_{i,j}},
\end{equation}
where $\zeta = e^{-2\lambda_{\textrm{B}}r_{\textrm{B}}^2}$ and $\eta_{\textrm{B}} = 2\lambda_{\textrm{B}}r_{\textrm{B}}(h_{\textrm{B}}-h_{\textrm{U}})/(h_{\textrm{A}}{-}h_{\textrm{U}})$~\cite{akramICCWS2020}.

\subsubsection{Throughput}

We assume that a user transmits data through its associated links only when the corresponding links are not blocked by dynamic blockers, since blockers are impenetrable.
We clarify that this assumption is adopted in previous relevant studies that investigated resource allocation in mmWave communication and THzCom environments where blockages exist~\cite{2018_TC_Idea,akramICC2020,MC1,MC4}.
\textcolor{black}{Considering this, the instantaneous achievable rate of the link between the $i$th user and the $j$th AP in sub-band $s$ is obtained as
\begin{align}\label{Equ:Rt}
R_{i,j,s}(t)=\begin{cases}
R_{i,j,s}^{\textrm{nb}}, & \textrm{with~the~probability~of}~p_{\textrm{nb}}(r_{i,j}),\\
R_{i,j,s}^{\textrm{b}}, & \textrm{with~the~probability~of}~(1-p_{\textrm{nb}}(r_{i,j})),
\end{cases} \forall i\in \mathcal {I},j\in \mathcal {J}, s\in \mathcal {S},
\end{align}
where $R_{i,j,s}^{\textrm{nb}}$ and $R_{i,j,s}^{\textrm{b}}$ are the achievable rates of the link between the $i$th user and the $j$th AP in sub-band $s$, when the link is not blocked and blocked by dynamic blockers, respectively, with $R_{i,j,s}^{\textrm{b}}=0$. Mathematically, $R_{i,j,s}^{\textrm{nb}}$ is written as~\cite{2011_Jornet_TWC}
\begin{align} \label{Equ:Rateijs1}
R_{i,j,s}^{\textrm{nb}}&=B_{s} \varphi  \log_{2}\left(1+\frac{P_{i,j,s}G_{\textrm{A}} G_{\textrm{U}}|\alpha_{i,j,s}|^2}{N_{0}B_{s}+\Psi_{i,j,s}}\right) ,   ~~~~~~ \forall i\in \mathcal {I},j\in \mathcal {J}, s\in \mathcal {S},
\end{align}}
where $P_{i,j,s}$ is the transmit power allocated by the  $i$th user for the link between itself and the $j$th AP in sub-band $s$ when the link is not blocked, $G_{\textrm{A}}$ and $G_{\textrm{U}}$ are the antenna gains at APs and users, respectively, $N_{0}$ is the noise spectral density, and $\Psi_{i,j,s}$ is the intra-band interference for the link between the $i$th user and the $j$th AP in sub-band $s$. Also, $|\alpha_{i,j,s}|^2$ is the path gain of the link between the $i$th user and the $j$th AP in sub-band $s$, and is given by
\begin{align} \label{Equ:Channel1}
|\alpha_{i,j,s}|^2=T_s \int_{f_s-\frac{B_s}{2}}^{f_s+\frac{B_s}{2}}  |H(f,d_{i,j})|^2 \textrm{d}f,~~~~~~ \forall i\in \mathcal {I},j\in \mathcal {J}, s\in \mathcal {S},
\end{align}
where $T_s$ is the duration of the pulse transmitted in sub-band $s$~\cite{2017N5}.
Note that we assume that the frame duration, defined as the time interval between two consecutive pulses, is higher than that of the duration of a pulse~\cite{2020_Chong_TWC_DistanceAdaptiveAbsorptionPeakModulation,HBM3}. This is to overcome the impact of pulse broadening in the time domain which may lead to inter-symbol interference~\cite{HBM3}.
This is reflected in \eqref{Equ:Rateijs1} using $\varphi$, where $\varphi$ is the ratio between the pulse duration and the frame duration.
Also, the impact of inter-band interference (IBI) is not considered in this work\footnote{Recently, IBI suppression schemes that can suppress IBI with minimal throughput degradation~\cite{HangNewChina} and waveform designs that minimize power leakages to adjacent bands~\cite{2020_Chong_TWC_DistanceAdaptiveAbsorptionPeakModulation}  were proposed.
Due to these and the fact there exist guard bands in between consecutive sub-bands in our considered spectrum allocation strategy, we clarify that,  it is reasonable to not consider the impact of IBI in this work.}~\cite{2017N5,2019_TC_JointUserAssociationandResourceAllocation_GeoffreyYeLi_subbandSC}.

We assume that all link distance values are available at the CCU~\cite{2017N5}.  Also, we consider that the CCU  determines the optimal  resource allocation policy to maximize a certain objective function.
We note that although transceivers are fixed, the blocked/non-blocked state of links can change very frequently over time due to the dynamic nature of blockers~\cite{MC1,akramICC2020}.
Therefore, it is challenging if resources are allocated dynamically over time based on the blocked/non-blocked state of links. Considering this, we assume that \textit{the CCU  determines the optimal resource allocation policy to maximize a certain long-term objective function\footnote{\textcolor{black}{We define the long-term value of a function as the average of the instantaneous values of this function over a relatively long-time duration. Specifically,  the long-term value of the function $\Theta$ is given by $\Theta=\lim_{T\rightarrow \infty}\frac{1}{T}\int_0^{T}\Theta(t)~\!\textrm{d}t$, where $\Theta(t)$ is the instantaneous value of the function $\Theta$ at time instant $t$.}}}.
Finally, considering that the CCU  synchronizes and combines the real-time data received from APs, the long-term throughput achieved by the $i$th user can be derived from \textcolor{black}{
\begin{align} \label{Equ:Rate0}
R_{i}&=\lim_{T\rightarrow \infty}\frac{1}{T}\int_0^{T} \sum _{j\in \mathcal {J}} \sum _{s\in \mathcal {S}}x_{i,j,s} R_{i,j,s}(t)~\!\textrm{d}t, ~~~\forall i\in \mathcal {I}.
\end{align}}
\textcolor{black}{We next substitute \eqref{Equ:Rt} in \eqref{Equ:Rate0} to obtain
\begin{align} \label{Equ:Rate1}
R_{i}&= \sum _{j\in \mathcal {J}} \sum _{s\in \mathcal {S}}x_{i,j,s} \lim_{T\rightarrow \infty}\frac{1}{T}\int_0^{T} R_{i,j,s}(t)~\!\textrm{d}t \notag \\
&= \sum _{j\in \mathcal {J}} \sum _{s\in \mathcal {S}}x_{i,j,s} \left( p_{\textrm{nb}}R_{i,j,s}^{\textrm{nb}} + \left(1-p_{\textrm{nb}}(r_{i,j})\right) R_{i,j,s}^{\textrm{b}} \right) \notag \\
&= \sum _{j\in \mathcal {J}} \sum _{s\in \mathcal {S}}x_{i,j,s}  p_{\textrm{nb}}R_{i,j,s}^{\textrm{nb}},
\end{align}}
where $p_{\textrm{nb}}$ and $ R_{i,j,s}^{\textrm{nb}}$ are given in \eqref{Equ:pLH} and \eqref{Equ:Rateijs1}, respectively, and $x_{i,j,s}$ is the user association and sub-band assignment indicator variable which will be discussed in the next section.

\section{Optimal Resource Allocation}
\label{Sec:ProbForm}

In this section, we first present the sub-band assignment strategy considered in this work. Thereafter, we present  the generalized resource allocation
problem. Finally, we discuss the challenges in solving the generalized resource allocation problem and discuss two modified problems which also represent two realistic considerations of the main optimization problem.

\subsection{Sub-band Assignment Strategy}
\label{Sec:Policy}

The limited advancement in THz band digital processors hinders the benefits of using novel but complex spectrum reuse schemes for enhancing the spectral efficiency of THzCom systems~\cite{HBM2}. Fortunately, it is possible to achieve high throughput in THzCom systems even with a lower spectral efficiency since the bandwidths available  at the THz band are in the order of hundreds of GHz~\cite{HBM3}. Thus, following~\cite{2017N5,HBM2,HBM3}, we assume that the associated links in the system are served by separate sub-bands. This assumption ensures intra-band interference-free data transmission, i.e., $\Psi_{i,j,s}=0$ in \eqref{Equ:Rateijs1}, thereby eliminating the hardware complexity and the signal processing overhead caused by frequency reuse in the system.
Under this assumption, we set the number of sub-bands to be equal to the number of associated links in the system, i.e., $S=I\times N$.

Let us define $x_{i,j,s}$ as the user association and sub-band assignment indicator variable, such that
\begin{align}\label{Equ:xijs}
x_{i,j,s}=\begin{cases}
1, & \textrm{if~sub-band}~s\textrm{~is~assigned~to~the~link~between~the~}i\textrm{th~user~and~the}~j\textrm{th~AP},\\
0, & \textrm{otherwise},
\end{cases}
\end{align}
where $ i\in \mathcal {I}, j\in \mathcal {J}$, and $s\in \mathcal {S}$. 
As each user associates with $N$ APs, we have
\begin{align}\label{Equ:xijsConst03}
\sum_{j\in \mathcal {J}}\sum_{s\in \mathcal {S}} x_{i,j,s}  = N, ~~\forall~i\in \mathcal {I}.
\end{align}
Also, as each associated link in the system is assigned with one sub-band, we have
\begin{align}\label{Equ:xijsConst01}
\sum_{s\in \mathcal {S}}x_{i,j,s}  = \begin{cases}
1, & \textrm{if~the~link~between~the~}i\textrm{th~user~and~the}~j\textrm{th~AP~is~an~associated~link},\\
0, & \textrm{otherwise},\end{cases}
\end{align}
where $i\in \mathcal {I}$ and $j\in \mathcal {J}$.
We note that \eqref{Equ:xijsConst01} can be equivalently written as
\begin{align}\label{Equ:xijsConst02}
\sum_{s\in \mathcal {S}}x_{i,j,s}  \leqslant 1, ~~~~~~ \forall i\in \mathcal {I},j\in \mathcal {J}.
\end{align}
Moreover, as each sub-band is assigned to one associated link, we have
\begin{align}\label{Equ:xijsConst04}
\sum_{i\in \mathcal {I}}\sum_{j\in \mathcal {J}}x_{i,j,s}  = 1, ~~~~~~ \forall s\in \mathcal {S}.
\end{align}

\begin{figure}[t]
\centering
   \centering
	\includegraphics[width=0.65\columnwidth]{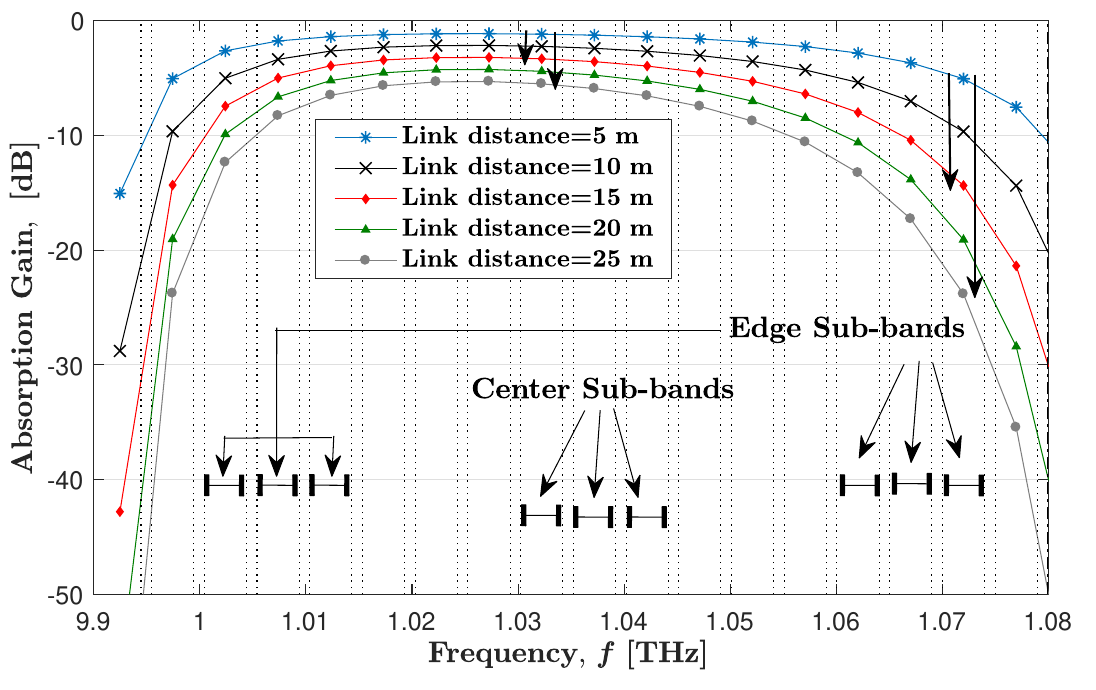}
	\vspace{-6mm}
	\caption{\textcolor{black}{Variation of absorption gain  for different link distances within the first THz TW above 1 THz.}}\label{Fig:DistVaryB}
\vspace{-3mm}
\end{figure}

We note that the frequency and distance-dependent nature of molecular absorption loss needs to be considered during user association and sub-band assignment in multiuser THzCom systems~\cite{HBM2,HBM3}. Specifically, as shown in Fig. \ref{Fig:DistVaryB}, on one hand, when the edge sub-bands (i.e., the sub-bands at the edges of the THz TW) are assigned to the links with different distances, the variation in molecular absorption gain is relatively high among the those links. On the other hand, when the center sub-bands (i.e., the sub-bands in the center region of the THz TW) are assigned to the links with different distances, the variation in molecular absorption gain is relatively low among the those links.
Due to these reasons, we introduce a constraint function given by
\begin{align} \label{Equ:AGConst}  
 |\alpha_{i,j,s}|^2\geqslant  x_{i,j,s} L_{\textrm{thr}},~~~~~~ \forall i\in \mathcal {I},j\in \mathcal {J}, s\in \mathcal {S},
\end{align}
where $L_{\textrm{thr}}$ is the path gain threshold imposed on the associated links. The constraint function in \eqref{Equ:AGConst} ensures that, while the associated links with longer distances can only be assigned to center sub-bands, the associated links with shorter distances can be assigned to any sub-band within the THz TW.

\subsection{Generalized Resource Allocation Problem}

It is of utmost importance to design novel and efficient spectrum allocation strategies to harness the potential of huge available bandwidths at the THz band. With this in mind, we study  generalized uplink resource allocation  for the considered MC-enabled multiuser THzCom system, with the primary focus on spectrum allocation. Specifically, we aim to maximize the long-term throughput under given sub-band bandwidth, power, and rate constraints.
This problem consists of sub-band assignment, user association identification, sub-band bandwidth allocation, and power control. Mathematically, this problem is formulated as

\begin{subequations}\label{OptProbOrg}
\begin{alignat}{2}
 \mathbf {P^{o}:}& \quad  \underset{\substack{x_{i,j,s},P_{i,j,s}, B_{s}\\ \forall i ,j, s}}{\textrm{maximize}}
& & \quad \mathcal {E}(R_{1} , R_{2} , \cdots,R_{I} )                                             \label{OptProbOrg-A} \\
&\quad    \textup{subject to}
 & &   \quad  \sum_{j\in \mathcal {J}} \sum_{s\in \mathcal {S}}   \zeta e^{-\eta_{\textrm{B}} r_{ij}} x_{i,j,s} P_{i,j,s} \leqslant P_{i}^{\textrm{max}}, ~~~~~~ \forall i\in \mathcal {I},                           \label{OptProbOrg-B} \\
& & &    \quad 0 \leqslant P_{i,j,s}\leqslant P_{i}^{\textrm{max}}, ~~~~~~ \forall i\in \mathcal {I},j\in \mathcal {J}, s\in \mathcal {S},                                     \label{OptProbOrg-B2} \\
& & &  \quad |\alpha_{i,j,s}|^2\geqslant  x_{i,j,s} L_{\textrm{thr}}, ~~~~~~ \forall i\in \mathcal {I},j\in \mathcal {J}, s\in \mathcal {S},                           \label{OptProbOrg1-C} \\
& & &  \quad R_{i,j,s}^{\textrm{nb}} \geqslant  x_{i,j,s}  R_{\textrm{thr}}, ~~~~~~ \forall i\in \mathcal {I},j\in \mathcal {J}, s\in \mathcal {S},                           \label{OptProbOrg-D}\\
& & &  \quad \sum_{s\in \mathcal {S}}B_{s}+(S-1)B_{\textrm{g}} = B_{\textrm{tot}},      \label{OptProbOrg-G} \\
& & &  \quad 0\leqslant B_{s} \leqslant B_{\textrm{max}},  ~~~~~~ \forall s\in \mathcal {S} ,                        \label{OptProbOrg-E}\\
%
& & &  \quad \sum_{j\in \mathcal {J}}\sum_{s\in \mathcal {S}} x_{i,j,s}  = N, ~~~~~~ \forall i\in \mathcal {I},                                                                                                  \label{OptProbOrg-I1}\\
& & &    \quad\sum_{s\in \mathcal {S}}x_{i,j,s}  \leqslant 1, ~~~~~~ \forall i\in \mathcal {I},j\in \mathcal {J},                                                 \label{OptProbOrg-H}\\
& & &  \quad \sum_{i\in \mathcal {I}}\sum_{j\in \mathcal {J}}x_{i,j,s}  = 1, ~~~~~~ \forall s\in \mathcal {S},                                                                                                  \label{OptProbOrg-I2}\\
& & &  \quad \sum_{i\in \mathcal {I}}\sum_{s\in \mathcal {S}}x_{i,j,s}  \leqslant M, ~~~~~~ \forall j\in \mathcal {J},                                                                                                  \label{OptProbOrg-I3}\\
& & &  \quad x_{i,j,s} \in \{0,1\} ,~~~~~~ \forall i\in \mathcal {I}, j\in \mathcal {J}, s\in \mathcal {S}.      \label{OptProbOrg-J}
\end{alignat}
\end{subequations}

\noindent In $ \mathbf {P^{o}}$, $\mathcal {E}(R_{1} , R_{2} , \cdots,R_{I} ) $ is the objective function imposed by the considered long-term throughput maximization strategy.
In this work, we aim to maximize the minimum throughput among all users, which has also been adopted by some of previous relevant studies, e.g.,~\cite{HBM2,2017N5}. Therefore, $\mathcal {E}(R_{1} , R_{2} , \cdots,R_{I} ) =\min_{i\in \mathcal {I}} \{ R_{i} \} $.
The justifications behind \eqref{OptProbOrg1-C}, \eqref{OptProbOrg-G}, \eqref{OptProbOrg-E}, \eqref{OptProbOrg-I1}, \eqref{OptProbOrg-H},   and \eqref{OptProbOrg-I2} are given in  \eqref{Equ:AGConst}, \eqref{Equ:BtotConst},  \eqref{Equ:BmaxConst}, \eqref{Equ:xijsConst03}, \eqref{Equ:xijsConst02},  and \eqref{Equ:xijsConst04}, respectively.
Moreover, \eqref{OptProbOrg-B} reflects the energy budget at each user. 
Specifically, recall that a user transmits data through its associated links only when the corresponding links are not blocked by dynamic blockers.
Thus, we have
$\sum_{j\in \mathcal {J}} \sum_{s\in \mathcal {S}}   x_{i,j,s} P_{i,j,s} T_{\textrm{tot}} p_{\textrm{nb}}(r_{i,j}) \leqslant E_{i}(T_{\textrm{tot}})$, $\forall i\in \mathcal {I}$, where $E_{i}(T_{\textrm{tot}})$ is the energy budget of the $i$th user for the duration $T_{\textrm{tot}}$. This leads to \eqref{OptProbOrg-B}, where $P_{i}^{\textrm{max}}= \frac{E_{i}(T_{\textrm{tot}})}{T_{\textrm{tot}}}$.
Furthermore, \eqref{OptProbOrg-D} ensures that  the achievable rate of all the associated links are lower bounded by the rate threshold $R_{\textrm{thr}}$. This guarantees that a user is in coverage as long as one of its associated links is in non-blocked state.
In addition, \eqref{OptProbOrg-I3} ensures that the maximum number of users with which each AP can associate is $M$.
Finally, \eqref{OptProbOrg-J} ensures that the user association and sub-band assignment indicator variable is binary, which reflects \eqref{Equ:xijs}.

The novelty of our considered generalized resource allocation problem, $\mathbf {P^{o}}$, is two-fold. First, we propose \textcolor{black}{multi-band-based} spectrum allocation with ASB, while to the best of our knowledge, recent studies in THz band \textcolor{black}{multi-band-based} spectrum allocation  have only considered ESB~\cite{HBM3,HBM2,2020_Chong_InfoCom_DABM,2019_Chong_DABM2,2021_THz_NOMA,2020ICC_NOMAforTHz,2017N5}.
Second, we  consider sub-band assignment in MC-enabled multiuser THzCom systems  under given path loss and rate constraints, while the previous relevant studies in THz band spectrum allocation   have considered  DAMC-based sub-band  assignment, the optimality of which for MC-enabled multiuser THzCom systems needs to be validated~\cite{HBM3,HBM2,2020_Chong_InfoCom_DABM,2019_Chong_DABM2}.
In Section \ref{Sec:Num}, we show the significance of these two considerations using numerical results.

\subsection{Modified Problems}
\label{Sec:Difficulty}

We note that it is challenging to solve the formulated optimization problem $\mathbf {P^{o}}$, due to the difficulty in obtaining the path gain, $|\alpha_{i,j,s}|^2$.
On one hand, obtaining $|\alpha_{i,j,s}|^2$ as per \eqref{Equ:Channel1} involves  an integral and the limits of the integral depend on the design variables $B_{k}$, $\forall$ $k \in \mathcal{S}$.
On the other hand, the values of $K(f)$ for all frequencies within the spectrum of interest, i.e., $K(f)$ for $f \in F$ where $F=\{f, f_{\textrm{ref}}-B_{\textrm{tot}}\leqslant f \leqslant f_{\textrm{ref}}\}$,  are required to obtain $|\alpha_{i,j,s}|^2$.
Although $K(f)$ for $f \in F$  can be found using the methodology in~\cite{2011_Jornet_TWC} with the aid of  HITRAN database values~\cite{2008_Hitran}, there does not exist a tractable expression  that maps $f$ to $K(f)$. This implies that it is extremely difficult, if not impossible, to analytically solve $\mathbf {P^{o}}$  using the traditional optimization theory techniques.
Therefore, we need to render the problem $\mathbf {P^{o}}$ computationally tractable. With this in mind,  we simplify some of the constraints in $\mathbf {P^{o}}$ to obtain two modified problems and solve them using  convex optimization~\cite{ConvexBoyedBook}.
We clarify that the two modified problems also represent two realistic scenarios of the considered THzCom system.
We next discuss the two modified problems as follows:

\subsubsection{Resource Allocation with ESB}

In this problem, we consider that all the sub-bands are of equal bandwidth.
We note that the analysis and the implementation of systems with ESB  is less complex than those with ASB, since the center frequency of sub-bands is known and fixed in the former.
Due to this, the spectrum arrangement with ESB has  been  widely considered in
other emerging THz band technologies and application scenarios, e.g., NOMA for THzCom systems~\cite{2019_Chong_DABM2,2021_THz_NOMA,2020ICC_NOMAforTHz}, THz band beamforming design~\cite{2020_WCNC_HangNan,HangNewChina} and THz backhauling~\cite{2020_Chong_InfoCom_DABM}.

We clarify that the novelty of  our resource allocation problem with ESB lies in the consideration of sub-band assignment in the considered MC-enabled multiuser THzCom system, while the previous relevant studies in THz band spectrum allocation   have considered  DAMC-based sub-band  assignment, the optimality of which for MC-enabled multiuser THzCom systems needs to be validated.
In Section \ref{Sec:P1}, we formulate the resource allocation problem with ESB for the considered THzCom system. Thereafter, we solve it using convex optimization.

\subsubsection{Resource Allocation with ASB in  either a PACSR or an NACSR}

\begin{figure}[t]
\centering
	\includegraphics[width=0.6\columnwidth]{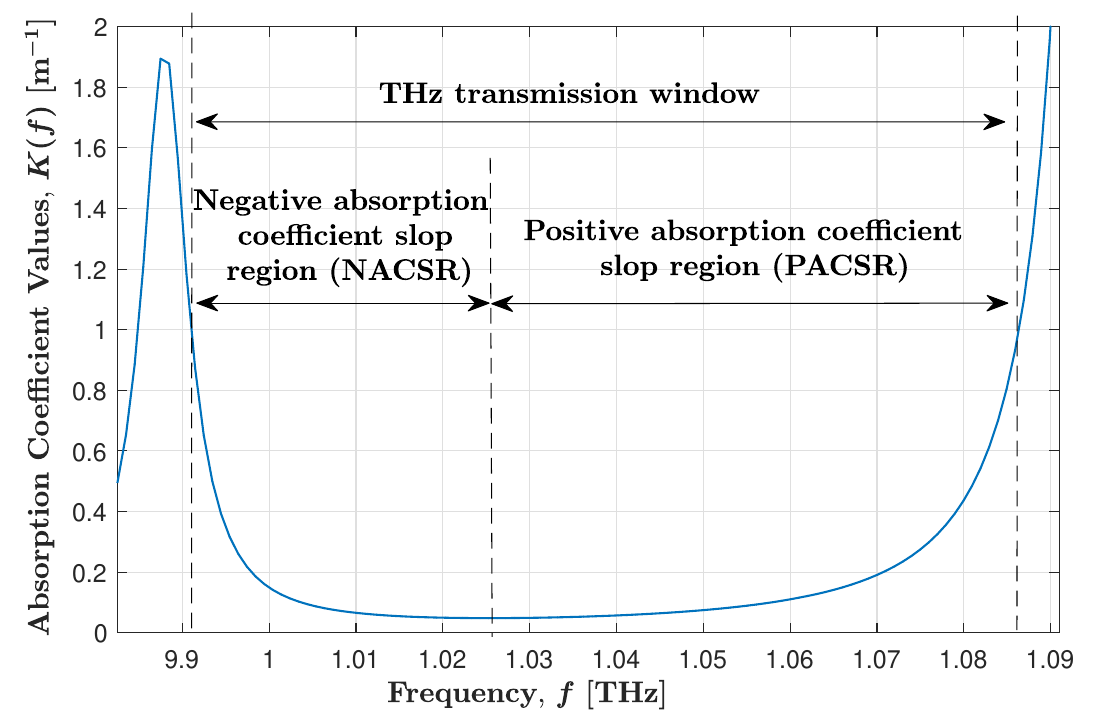}
	\vspace{-6mm}
	\caption{Illustration of the positive and negative absorption coefficient slope regions within a THz TW.}\label{Fig:VaryingK} 
\end{figure}

In this problem, we consider that the spectrum of interest fully exists either in a PACSR or  an NACSR of the THz band. 
We note that PACSRs and NACSRs are defined as the regions  within the THz TWs where $\frac{\partial K(f)}{\partial f}> 0$ and $\frac{\partial K(f)}{\partial f}< 0$, respectively, as depicted in  Fig. \ref{Fig:VaryingK}.
We clarify that it is reasonable to focus on the allocation of the spectrum that exists either within a PACSR or an NACSR, since the available bandwidths in each PACSR and NACSR at the THz band are in the order of  tens of GHz, as shown in Table \ref{tab:Bandwidth}.

We clarify that the main novelty of our resource allocation problem with ASB in a PACSR lies in the consideration of \textcolor{black}{multi-band-based} spectrum allocation with ASB~\cite{HBM3,HBM2,2020_Chong_InfoCom_DABM,2019_Chong_DABM2,2021_THz_NOMA,2020ICC_NOMAforTHz,2017N5}.
In Section \ref{Sec:P2},    by modifying $\mathbf {P^{o}}$ we first formulate resource allocation problem with ASB in a PACSR  for the considered THzCom system. Thereafter, with reasonable approximations, we transform the problem into a  mixed-integer nonlinear problem. Finally, we transform the problem into a  convex program utilizing  additional  transformations and then solve it using  convex optimization.
It is noted that the formulation and the solution of the resource allocation problem with ASB in an NACSR are similar to those in a PACSR.
Hence, the resource allocation problem with ASB in an NACSR can be formulated and solved by directly modifying the problem formulation and the solution presented for PACSR in Section$~$\ref{Sec:P2}, which is omitted in this work.


\begin{table}[!t]
\caption{Available bandwidths within PACSR and NACSR at the THz band\protect\footnote{XXX}~\cite{2011_Jornet_TWC}.}\vspace{-7mm}
\begin{center}
\begin{tabular}{|c|c|c|c|}
\hline
\multicolumn{2}{|c|}{\textbf{PACSR}} &\multicolumn{2}{|c|}{\textbf{NACSR}} \\
\hline
\textbf{Frequency Range} $\mathrm{\mathbf{(THz)}}$& \textbf{Bandwidth} $\mathrm{\mathbf{(GHz)}}$ & \textbf{Frequency Range} $\mathrm{\mathbf{(THz)}}$ & \textbf{Bandwidth} $\mathrm{\mathbf{(GHz)}}$         \\
\hline
$0.4934-0.5521$  & $58.7$ &  $0.5620-0.6107$  & $48.7$\\
\hline
 $0.6734-0.7487$ & $78.5$ &  $0.6207-0.6515$  & $30.8$\\
\hline
 $0.8702-0.9159$ & $45.7$ &  $0.7559-0.8513$  & $95.4$\\
 \hline
 $1.0250-1.0870$  & $62.0$ & $0.9905-1.025$  & $34.5$\\
\hline
\end{tabular}\label{tab:Bandwidth}
\end{center}\vspace{-8mm}
\end{table}
\footnotetext{Values are calculated  for the standard atmosphere with $10\%$ humidity, while considering that the frequencies where absorption coefficient values are less that one, belongs to THz TWs.}

\label{Sec:P2}

\section{Resource Allocation with ESB}
\label{Sec:P1}

In this section, we focus on \emph{the resource allocation problem with ESB} for the MC-enabled multiuser THzCom system. We first formulate the problem and then solve it using  convex optimization.

\subsection{Problem Reformulation}

When all the sub-bands are of equal bandwidth,  the bandwidth of sub-bands, $B_{s}$, which is a design variable in $ \mathbf {P^{o}}$, can be obtained  by using \eqref{Equ:BtotConst} as $B_{s}=\left(\digamma_{\!\!s}+\frac{1}{2}(1-2 s)B_s \right)/S$.
This helps to  solve $ \mathbf {P^{o}}$ in a tractable manner, since $|\alpha_{i,j,s}|^2$  $\forall$ $i\in \mathcal {I},j\in \mathcal {J}, s\in \mathcal {S}$ can be calculated from \eqref{Equ:Channel1} by using the known values of $B_{k}$, where $k \in \mathcal{S}$. 
Considering this,  the resource allocation problem with ESB for the MC-enabled multiuser THzCom system, $ \mathbf {P^{o}_{1}}$, is formulated as a  computationally tractable optimization program, given by

\begin{alignat}{2}
 \mathbf {P^{o}_{1}:} \quad &  \underset{\substack{x_{i,j,s},P_{i,j,s}\\ \forall i ,j, s}}{\textrm{minimize}}
& & \quad \underset {i\in \mathcal {I}}{\max }\{- R_{i}\}                                              \notag\\
&  \textup{subject to}
& &    \quad  \eqref{OptProbOrg-B}-\eqref{OptProbOrg-D}, \eqref{OptProbOrg-H}-\eqref{OptProbOrg-J} . \label{OptProb1}
\end{alignat}
The optimization problem $\mathbf {P^{o}_{1}}$ yields the optimal sub-band assignment and transmit power for the ESB scenario. In the next subsection, we present the solution adopted to solve $\mathbf {P^{o}_{1}}$.

\subsection{Problem Solution}
\label{Sec:P1Sol}
The formulated problem $\mathbf {P^{o}_{1}}$ is a mixed-integer nonlinear problem, which is non-convex in its original form~\cite{ConvexBoyedBook}.
In particular, the non-convexity arises from the binary constraint function \eqref{OptProbOrg-J} since it spans a disjoint feasible solution set. Therefore, it is challenging to determine the globally optimal solution to $\mathbf {P^{o}_{1}}$ problem.
To overcome this obstacle,
we transform the binary variables in $\mathbf {P^{o}_{1}}$ into real variables~\cite{2014_PenaltyFactor2,BinaryTrans1,BinaryTrans2,Sheeraz2020}. In doing so, we rewrite  the binary constraint function  \eqref{OptProbOrg-J} equivalently as the combination of two constraints, given by
\begin{alignat}{2}
& 0 \leqslant x_{i,j,s} \leqslant 1 ,~~~~~~ \forall  i\in \mathcal {I}, j\in \mathcal {J}, s\in \mathcal {S},   \label{OptProb1Sol1-B}
\end{alignat}
and
\begin{alignat}{2}
& \sum_{i\in \mathcal {I}} \sum_{j\in \mathcal {J}} \sum_{s\in \mathcal {S}} \Big ( x_{i,j,s} {-} x^2_{i,j,s} \Big) \leqslant 0 ~. \label{Equ:TransBin12}
\end{alignat}
\textcolor{black}{Considering \eqref{OptProb1Sol1-B} and \eqref{Equ:TransBin12}, we transform $\mathbf {P^{o}_{1}}$ into the following equivalent problem, given by
\begin{alignat}{3} \label{OptProb1SolX}
\mathbf {\ddot{P}^{o}_{1}:} & \quad \underset{\substack{x_{i,j,s},P_{i,j,s}\\ \forall i ,j, s}}{\textrm{minimize}}
& & \quad \quad \underset {i\in \mathcal {I}}{\max }\{- R_{i}\}   \notag                                 \\
& \quad \textup{subject to}
& &    \quad  \eqref{OptProbOrg-B}-\eqref{OptProbOrg-D}, \eqref{OptProbOrg-H}-\eqref{OptProbOrg-I3} , \eqref{OptProb1Sol1-B}, \eqref{Equ:TransBin12},
\end{alignat}
\noindent We note that   in $\mathbf {\ddot{P}^{o}_{1}}$, although $x_{i,j,s}$ is relaxed to be real between zero and one, the constraint \eqref{Equ:TransBin12} guarantees that $x_{i,j,s}$ can only be zero or one, since $x_{i,j,s} {-} x^2_{i,j,s}\geq 0, \forall  i\in \mathcal {I}, j\in \mathcal {J}, s\in \mathcal {S}$. Next, considering the practical computational feasibility, we relax the constraint in \eqref{Equ:TransBin12} and include it as a penalty function in  the objective function. In doing so, we transform $\mathbf {\ddot{P}^{o}_{1}}$ into the following problem, given by
\begin{alignat}{3} \label{OptProb1Sol}
\mathbf {\bar{P}^{o}_{1}:} & \quad \underset{\substack{x_{i,j,s},P_{i,j,s}\\ \forall i ,j, s}}{\textrm{minimize}}
& & \quad \quad \underset {i\in \mathcal {I}}{\max }\{- R_{i}\} +\Lambda_1 \sum_{i\in \mathcal {I}} \sum_{j\in \mathcal {J}} \sum_{s\in \mathcal {S}} \Big ( x_{i,j,s} {-} x^2_{i,j,s} \Big)     \notag                                 \\
& \quad \textup{subject to}
& &    \quad  \eqref{OptProbOrg-B}-\eqref{OptProbOrg-D}, \eqref{OptProbOrg-H}-\eqref{OptProbOrg-I3} , \eqref{OptProb1Sol1-B},
\end{alignat}
where $\Lambda_1 \geqslant 0$ is the constant penalty factor and $\sum_{i\in \mathcal {I}} \sum_{j\in \mathcal {J}} \sum_{s\in \mathcal {S}} \Big ( x_{i,j,s} {-} x^2_{i,j,s} \Big)$ is the penalty function on the violation of the binary constraint over the objective function.
We clarify that,  for  the optimization problem $\mathbf {\bar{P}^{o}_{1}}$,  different optimal values can be obtained by varying $\Lambda_1$.
On one hand, by setting $\Lambda_1$ extremely high, the optimal values of  $\mathbf {\bar{P}^{o}_{1}}$ can be obtained, which would guarantee the penalty function to be extremely small, which would in turn lead to negligible violation of the binary constraint. On the other hand, by setting $\Lambda_1$ extremely low, the optimal values of  $\mathbf {\bar{P}^{o}_{1}}$ can be obtained, which would be close to the optimal value of $\mathbf {\ddot{P}^{o}_{1}}$, at the cost of high violation of the binary constraint.
 Thus, $\Lambda_1$ must be carefully selected such that the optimal value of  $\mathbf {\bar{P}^{o}_{1}}$ is reasonably close to the optimal value of $\mathbf {\ddot{P}^{o}_{1}}$, and the violation of the binary constraint is not high.   Considering this, we choose $\Lambda_1$ to be sufficiently large and introduce a numerical tolerance level on the penalty function such that it is acceptable to have $\sum_{i\in \mathcal {I}} \sum_{j\in \mathcal {J}} \sum_{s\in \mathcal {S}} \Big ( x_{i,j,s} {-} x^2_{i,j,s} \Big) < \epsilon$, where $\epsilon$ is a very small positive number~\cite{2014_PenaltyFactor2,BinaryTrans1,BinaryTrans2}.}

After this manipulation,
we note that the penalty function in $\mathbf {\bar{P}^{o}_{1}}$  is non-convex in $x_{i,j,s}$.
To handle this non-convexity, we consider the Taylor approximation of the function $g(x)=x-x^2$~\cite{Sheeraz2020}. We observe that the Taylor approximation of $g(x)$ is convex and it provides an upper bound on $g(x)$. Considering this, we obtain the convex upper bound on the penalty function in \eqref{OptProb1Sol}, given by
\begin{align}
&\sum_{i\in \mathcal {I}} \sum_{j\in \mathcal {J}} \sum_{s\in \mathcal {S}} \Big (x_{i,j,s} \big (1 {-} 2 x^{(\kappa)}_{i,j,s} \big) {+} \big (x^{(\kappa)}_{i,j,s} \big)^{2} \Big)   \geqslant \sum_{i\in \mathcal {I}} \sum_{j\in \mathcal {J}} \sum_{s\in \mathcal {S}} \Big ( x_{i,j,s} {-} x^2_{i,j,s} \Big) . \label{Equ:TransBin13}
\end{align}
Finally, using \eqref{Equ:TransBin13} in \eqref{OptProb1Sol}, we transform $\mathbf {P^{o}_{1}}$ into the following approximate convex problem, given by
\begin{alignat}{2}
\mathbf {\hat{P}^{o}_{1}:}\quad &  \underset{\substack{x_{i,j,s},P_{i,j,s}\\ \forall i ,j, s}}{\textrm{minimize}}
& & \quad \underset {i\in \mathcal {I}}{\max }\{- R_{i}\}  +\Lambda_1 \sum_{i\in \mathcal {I}} \sum_{j\in \mathcal {J}} \sum_{s\in \mathcal {S}} \Big (x_{i,j,s} \big (1 {-} 2 x^{(\kappa)}_{i,j,s} \big) {+} \big (x^{(\kappa)}_{i,j,s} \big)^{2} \Big)                                            \notag \\
&  \textup{subject to}
& &  \quad \eqref{OptProbOrg-B}-\eqref{OptProbOrg-D}, \eqref{OptProbOrg-H}-\eqref{OptProbOrg-I3},\eqref{OptProb1Sol1-B}. \label{OptProb1Sol2}   
\end{alignat}

We clarify that $\mathbf {\hat{P}^{o}_{1}}$ can be solved efficiently by using standard convex problem solvers, such as CVX~\cite{CVX}.
However, the optimal value of $\mathbf {\hat{P}^{o}_{1}}$ is a global upper bound on the  optimal value of $\mathbf {P^{o}_{1}}$, since the upper bound on the penalty function is utilized in $\mathbf {\hat{P}^{o}_{1}}$.
Therefore, by using the successive convex approximation
(SCA) technique, we propose an iterative algorithm to tighten
the upper bound obtained from solving $\mathbf {\hat{P}^{o}_{1}}$, which is
summarized in Algorithm \ref{Alg:Alg1}.

We clarify that $\mathbf {\hat{P}^{o}_{1}}$  is solved in each iteration of Algorithm \ref{Alg:Alg1} with a polynomial computational complexity.
Also, there are totally $(2IJS)$ decision variables and $ (3IJS+2IJ+2I+S+J)$ convex constraints in the convex problem $\mathbf {\hat{P}^{o}_{1}}$.
Thus, solving $\mathbf {\hat{P}^{o}_{1}}$  requires a complexity of
$\mathcal{O} \big( (2IJS)^3(3IJS+2IJ+2I+S+J) \big)$
\cite{2020_TC_ClusterBasedResourceAllocationNPProbWith,2014_TWC_OptimaztionFramework3}.

\section{Resource Allocation with ASB in a PACSR}

\label{Sec:P2}

In this section, we focus on \emph{the resource allocation problem with ASB in  a PACSR}  for the MC-enabled multiuser THzCom system. We first formulate the problem and then solve it using  convex optimization.
\subsection{Problem Reformulation}

By modifying $\mathbf {P^{o}}$, the resource allocation problem with ASB in a PACSR for the MC-enabled multiuser THzCom system is formulated as
\begin{subequations}\label{OptProbOrg2}
\begin{alignat}{2}
\mathbf {P^{o}_2:}\quad &  \underset{\substack{x_{i,j,s},P_{i,j,s}, B_{s}\\ \forall i ,j, s}}{\textrm{minimize}}
& & \quad \underset {i\in \mathcal {I}}{\max }\{- R_{i}\}                                           \label{OptProbOrg2-A}  \\
&  \textup{subject to}
& &  \quad \eqref{OptProbOrg-B} - \eqref{OptProbOrg-D}, \eqref{OptProbOrg-E},
\eqref{OptProbOrg-H}-\eqref{OptProbOrg-J}    ,        \notag   \\
& & &  \quad \sum_{s\in \mathcal {S}}B_{s} = \bar{B}_{\textrm{tot}},      \label{OptProbOrg2-D}
%
\end{alignat}
\end{subequations}
where $\bar{B}_{\textrm{tot}}=B_{\textrm{tot}}-(S-1)B_{\textrm{g}}$. The optimization problem $\mathbf {P^{o}_2}$ yields the optimal sub-band assignment, sub-band bandwidth, and the  and transmit power for the scenario when the  spectrum of interest that is to be allocated exists in a PACSR of the THz TW.

\begin{algorithm}[t]

\caption{Iterative Approach for Resource Allocation with ESB}
\begin{algorithmic}[1] \label{Alg:Alg1}
\STATE \textbf{Initialization}: Set iteration count $\kappa=0$. Set initial point for $x^{ (\kappa)}_{i,j,s}=0.5, \, \forall i\in \mathcal {I}, j\in \mathcal {J}, s\in \mathcal {S}$.  Select a reasonably high penalty factor $\Lambda_2$ and low tolerance value $\epsilon$.
\WHILE{$\sum\limits_{i\in \mathcal {I}} \sum\limits_{j\in \mathcal {J}} \sum\limits_{s\in \mathcal {S}} \Big (x_{i,j,s} \big (1 {-} 2 x^{(\kappa)}_{i,j,s} \big) {+} \big (x^{(\kappa)}_{i,j,s} \big)^{2} \Big)  \geqslant \epsilon$}
   \STATE Solve  \eqref{OptProb1Sol2} using point $x^{(\kappa)}_{i,j,s}, \, \textrm{where}~ i\in \mathcal {I}, j\in \mathcal {J}, s\in \mathcal {S}$ and obtain solution parameters $P_{i,j,s}^{*}, x_{i,j,s}^{*}, \forall i\in \mathcal {I}, j\in \mathcal {J}, s\in \mathcal {S}$.
   \STATE Update point $x^{(\kappa+1)}_{i,j,s} = x_{i,j,s}^*,  \forall i\in \mathcal {I}, j\in \mathcal {J}, s\in \mathcal {S}$.
   \STATE Update iteration count $\kappa = \kappa + 1$.
\ENDWHILE
\end{algorithmic}
\end{algorithm}

We note that the difficulties related to obtaining $|\alpha_{i,j,s}|^2$, that were mentioned in Section \ref{Sec:Difficulty}, still exist when solving $\mathbf {P^{o}_2}$. To overcome this, we make two assumptions which enable us to obtain $|\alpha_{i,j,s}|^2$ using a tractable expression of the design variables  $B_{k}$, $\forall$ $k \in \mathcal{S}$.
We next present these assumptions as follows:

First, we overcome the obstacle caused due to $|\alpha_{i,j,s}|^2$ being obtained as in \eqref{Equ:Channel1} using an integral.
We observe that although the molecular absorption coefficient is frequency-dependent, its variation within sub-bands are relative small when sub-bands exist within THz TWs. Considering this, for tractable analysis, we represent $|\alpha_{i,j,s}|^2$ assuming that the molecular absorption coefficient remains unchanged within each sub-band\footnote{We clarify that the assumption that the molecular absorption coefficient remains unchanged within the bandwidth of interest  has been adopted in previous THz band studies for tractable system design such as~\cite{akram2020JSAC,HBM2,2020_WCL_Multi_RATMCforTHz}.}~\cite{akram2020JSAC,HBM2,2020_WCL_Multi_RATMCforTHz}.
This consideration enables us to obtain $|\alpha_{i,j,s}|^2$ as 
\begin{align} \label{Equ:Channelapprox1}
|\alpha_{i,j,s}|^2 &=  \frac{\varrho }{f^2_s d^2_{i,j}}e^{-K(f_s) d_{i,j}},
\end{align}
where $\varrho\;\triangleq \left(\frac{c}{4 \pi }\right)^{2}$.

Second, we tackle the intractability caused due to the lack of tractable expression for $K(f)$. To this end, we first obtain  the values of $K(f)$ for all the values of $f$ in the spectrum of interest using the HITRAN database and observe its variation within the spectrum of interest~\cite{2008_Hitran}. We notice that the variation  of $K(f)$ in spectra that exist within PACSRs generally exhibits the behavior of an exponential function of $f$ (see Fig. \ref{Fig:VaryingK}).  Therefore, through curve fitting,
we model $K(f)$ in the spectrum of interest  using an exponential function of $f$, given by
\begin{equation}\label{Equ:CurveFit}
  \hat{K}(f)=e^{\sigma_{1} +\sigma_{2} f}+\sigma_{3},
\end{equation}
where  $\hat{K}(f)$ is the approximated molecular absorption coefficient at $f$ and  $\{\sigma_{1},\sigma_{2},\sigma_{3}\}$ are the model parameters obtained  with $\varepsilon-$confidence interval for the spectrum of interest.
We note that the values of $\hat{K}(f)$ in other spectra that exist within PACSRs can also be models using exponential functions of $f$. However, the model parameters would differ from one spectrum to another.

We next substitute \eqref{Equ:f_s} and \eqref{Equ:CurveFit}  in \eqref{Equ:Channelapprox1} to obtain 
\begin{equation}\label{Equ:AGF}
  |\alpha_{i,j,s}|^2= \varrho ~\! \frac{ e^{-\hat{K}(f_s) d_{i,j}}}{\left(f_s d_{i,j} \right)^2}= \varrho  ~\! \frac{e^{- d_{i,j}[e^{\sigma_{1} +\sigma_{2} (\digamma_{\!\!s} - \sum_{k\in \mathcal {S}}a_{s,k}B_{k}) }+\sigma_{3}]}}{(\digamma_{\!\!s} - \sum_{k\in \mathcal {S}}a_{s,k}B_{k})^2 d_{i,j}^{2}},
\end{equation}
where  $\digamma_{\!\!s}=f_{\textrm{ref}}-(s-1)B_{\textrm{g}}$ and $a_{s,k}=1~\textrm{if}~s >k$, $a_{s,k}=\frac{1}{2}~\textrm{if}~s =k$, and $a_{s,k}=0~\textrm{otherwise}$.
Thereafter, using \eqref{Equ:AGF},  the multiuser optimization problem for the PACSR, $\mathbf {P^{o}_2}$,  is approximated as a mixed-integer nonlinear problem, given by
\begin{alignat}{2}
\mathbf {\bar{P}^{o}_2:}\quad &  \underset{\substack{x_{i,j,s},P_{i,j,s}, B_{s}\\ \forall i ,j, s}}{\textrm{minimize}}
& & \quad \underset {i\in \mathcal {I}}{\max }\{- R_{i}\}   \notag                                         \\
&  \textup{subject to}
& &  \quad \eqref{OptProbOrg-B} - \eqref{OptProbOrg-D},
 \eqref{OptProbOrg-E},
\eqref{OptProbOrg-H} - \eqref{OptProbOrg-J},\eqref{OptProbOrg2-D}.  \label{OptProbOrg3}
\end{alignat}

\subsection{Problem Solution}

The formulated problem $\mathbf {\bar{P}^{o}_2}$ is a non-convex problem.
In particular, the non-convexity in $\mathbf {\bar{P}^{o}_2}$ arises due to three reasons. First, $R_{i,j,s}^{\textrm{nb}}$ appearing in the objective function in \eqref{OptProbOrg3} and the constraint function \eqref{OptProbOrg-D} is not differentiable at $B_s=0$. Second, the objective function in \eqref{OptProbOrg3} and the constraint  functions \eqref{OptProbOrg1-C} and \eqref{OptProbOrg-D} are non-convex  w.r.t. the design variable $B_{\nu}$, $\forall$ $\nu \in \mathcal{S}$. This is due to the fact that $|\alpha_{i,j,s}|^2$  and
$R_{i,j,s}^{\textrm{nb}}$  are not concave  w.r.t. the design variable $B_{\nu}$, $\forall$ $\nu \in \mathcal{S}$. Third,
the design variable $x_{i,j,s}$ is binary.
We next tackle these challenges as follows:

We first handle the non-differentiability of $R_{i,j,s}^{\textrm{nb}}$ at $B_s=0$.
When the objective function is differentiable in an open domain, the Karush-Kuhn-Tucker (K.K.T.) conditions are sufficient and necessary for the optimal solution~\cite{ConvexBoyedBook}.
However, we note that $R_{i,j,s}^{\textrm{nb}}$ in $\mathbf {\bar{P}^{o}_2}$ is not differentiable at $B_s=0$.  Thus, to utilize the K.K.T. conditions to characterize the optimality of the problem in $\mathbf {\bar{P}^{o}_2}$, we consider the following approximate problem, given by
\begin{subequations}\label{OptProb2Sol1}
\begin{alignat}{2}
\mathbf {\bar{P}^{o}_2(\delta):} \quad &  \underset{\substack{x_{i,j,s},P_{i,j,s}, B_{s}\\ \forall i ,j, s}}{\textrm{minimize}}
& & \quad \underset {i\in \mathcal {I}}{\max }\{- R_{i}\}              \\
&  \textup{subject to}
& &  \quad  \eqref{OptProbOrg-B} - \eqref{OptProbOrg-D},
\eqref{OptProbOrg-H} - \eqref{OptProbOrg-J} ,\eqref{OptProbOrg2-D},  \notag  \\
& & &  \quad \delta \leqslant B_{s} \leqslant B_{\textrm{max},s}, ~~~~~~ \forall s\in \mathcal {S}, \label{OptProb2Sol1-B}
\end{alignat}
\end{subequations}
where $\delta$ is a very small positive number. We note that the optimal value of $\mathbf {\bar{P}^{o}_2(\delta)}$ converges to the optimal value of $\mathbf {\bar{P}^{o}_2}$ when $\delta\rightarrow0^+$, i.e.,
$\lim_{\delta\rightarrow0^+} \mathbf {\bar{P}^{o}_2(\delta)}=\mathbf {\bar{P}^{o}_2}$~\cite{2015_TC_BA2_forBminArguement}.

We next handle the non-concavity of $|\alpha_{i,j,s}|^2$  and $R_{i,j,s}^{\textrm{nb}}$  w.r.t. the design variable $B_{\nu}$, $\forall$ $\nu \in \mathcal{S}$.
For this purpose, we consider the following substitution for $B_{s}$, given by
\begin{equation}\label{Equ:Trans1}
  B_{s}=\xi_s+\omega_s\log(\varsigma_s Z_{s}), ~~s\in \mathcal {S},
\end{equation}
where $\xi_{s},\omega_{s},\varsigma_{s}>0$ $\forall$ $\nu \in \mathcal{S}$.
Thereafter, substituting \eqref{Equ:Trans1} in \eqref{Equ:AGF} and \eqref{Equ:Rateijs1}, we obtain
\begin{equation}\label{Equ:AG21}
  |\alpha_{i,j,s}|^2=\varrho ~\!\frac{e^{- d_{i,j}[e^{\sigma_{1} +\sigma_{2} (\digamma_{s}- \sum_{k\in \mathcal {S}}a_{s,k}(\xi_{k}+\omega_k\log(\varsigma_{k} Z_{k})))}+\sigma_{3}]}}{(\digamma_{s}- \sum_{k\in \mathcal {S}}a_{s,k}(\xi_{k}+\omega_{k}\log(\varsigma_{k} Z_{k})))^{2}d_{i,j}^{2}}, ~~\forall i\in \mathcal {I},j\in \mathcal {J}, s\in \mathcal {S}
\end{equation}
and
\begin{align}\label{Equ:Rateijs22}
  R_{i,j,s}^{\textrm{nb}}&=(\xi_s+\omega_s\log(\varsigma_s Z_{s}))  \varphi \log_{2}\left(1+\frac{P_{i,j,s}G_{\textrm{A}} G_{\textrm{U}}|\alpha_{i,j,s}|^2}{ N_{0}(\xi_s+\omega_s\log(\varsigma_s Z_{s})) }\right), ~~ \forall i\in \mathcal {I},j\in \mathcal {J}, s\in \mathcal {S} .
\end{align}
We observe that the expression for $|\alpha_{i,j,s}|^2$ and $R_{i,j,s}^{\textrm{nb}}$ in \eqref{Equ:AG21} and \eqref{Equ:Rateijs22}, respectively, are highly non-linear w.r.t. $Z_{\nu}$ for ${\nu}\in \mathcal {S}$. However, through careful deliberation,  we arrive at the following Lemma.
\begin{lemma} \label{Lem:pLH}
It is found that $|\alpha_{i,j,s}|^2$ and $R_{i,j,s}^{\textrm{nb}}$ in \eqref{Equ:AG21}
and \eqref{Equ:Rateijs22}, respectively, are concave w.r.t. $Z_{\nu}$ $\forall$ $i\in\mathcal{I},j\in \mathcal {J}, s\in \mathcal {S},{\nu}\in \mathcal {S}$ when $1/\omega_{\nu}>\bar{\omega}$, where $\bar{\omega}=\sigma_{2}\left(D\hat{K}(f_{\textrm{ref}})e^{D\sigma_3}-1\right)$, $D>d_{\textrm{max}}$, $d_{\textrm{max}}$ is the maximum of link distances, i.e., $d_{\textrm{max}}=\underset {i\in \mathcal {I},j\in \mathcal {J}}{\max }\{ d_{i,j}\}$.

\textit{~~~Proof:} See Appendix \ref{app:Derive_Lemma1}.
 \hfill $\blacksquare$
\end{lemma}
Following Lemma 1, we select $\omega_{\nu}$ such that $1/\omega_{\nu}>\bar{\omega}$ $\forall$ ${\nu}\in \mathcal {S}$ to ensure the convexity of the objective function in \eqref{OptProbOrg3} and constraint functions \eqref{OptProbOrg1-C} and \eqref{OptProbOrg-D} w.r.t. $Z_{\nu}$. Thereafter,
we transform $\mathbf {\bar{P}^{o}_2(\delta)}$ into the following equivalent problem, given by
\begin{subequations} \label{P2V3}
\begin{alignat}{2}
 &  \underset{\substack{x_{i,j,s},P_{i,j,s}, Z_{s}\\ \forall i ,j, s}}{\textrm{minimize}}
& & \quad \underset {i\in \mathcal {I}}{\max }\{- R_{i}\}   \\
&  \textup{subject to}
& &  \quad\eqref{OptProbOrg-B} - \eqref{OptProbOrg-D},
\eqref{OptProbOrg-H} - \eqref{OptProbOrg-J} ,    \notag          \\
& & &  \quad \prod_{s=1}^{S}Z_{s}^{\omega_{s}}- \prod_{s=1}^{S} \varsigma_{s}^{-\omega_{s}}e^{\bar{B}_{\textrm{tot}}-\sum_{s\in \mathcal {S}} \xi_{s}} \leqslant 0,      \label{OptProb2Sol2-C}  \\
& & &  \quad Z_{\textrm{min},s} \leqslant Z_{s} \leqslant Z_{\textrm{max},s} , ~~~~~~ \forall s\in \mathcal {S},   \label{OptProb2Sol2-A}
\end{alignat}
\end{subequations}
where  $ Z_{\textrm{min},s}=\frac{1}{\varsigma_s}e^{\frac{\delta-\xi_s}{\omega_s}}$  and $ Z_{\textrm{max},s}=\frac{1}{\varsigma_s}e^{\frac{B_{\textrm{max}}-\xi_s}{\omega_s}}$.
We note that \eqref{OptProb2Sol2-C} and \eqref{OptProb2Sol2-A}  are obtained by substituting \eqref{Equ:Trans1} in \eqref{OptProbOrg2-D}
 and \eqref{OptProb2Sol1-B}, respectively.

We finally handle the non-convexity arising in \eqref{P2V3} from the binary constraint function \eqref{OptProbOrg-D}. To this end, we utilize the same procedure adopted in Section \ref{Sec:P1Sol}, where the binary variables are transformed into real variables~\cite{BinaryTrans1,BinaryTrans2}.  In doing so, we approximate \eqref{P2V3} into the following
convex problem, given by
\begin{alignat}{2}
 \mathbf {\hat{P}^{o}_{2}:}\quad &  \underset{\substack{x_{i,j,s},P_{i,j,s}, Z_{s}\\ \forall i ,j, s}}{\textrm{minimize}}
& & \quad \underset {i\in \mathcal {I}}{\max }\{- R_{i}\}   +\Lambda_2 \sum_{i\in \mathcal {I}} \sum_{j\in \mathcal {J}} \sum_{s\in \mathcal {S}}  \Big (x_{i,j,s} \big (1 {-} 2 x^{(\kappa)}_{i,j,s} \big) {+} \big (x^{(\kappa)}_{i,j,s} \big)^{2} \Big)                                            \notag     \\
&  \textup{subject to}
& &  \quad\eqref{OptProbOrg-B} - \eqref{OptProbOrg-D},
\eqref{OptProbOrg-H} - \eqref{OptProbOrg-I3} , \eqref{OptProb2Sol2-A}-\eqref{OptProb2Sol2-C},    \notag       \\
& & &  \quad 0 \leqslant x_{i,j,s} \leqslant 1 ,~~~~~~ \forall i\in \mathcal {I}, j\in \mathcal {J}, s\in \mathcal {S},      \label{OptProb2Sol}
\end{alignat}
where $\Lambda_2 \geqslant 0$ is the constant penalty factor. We note that, similar to the observations in Section \ref{Sec:P1}, the optimal value of $\mathbf {\hat{P}^{o}_{2}}$ is a global upper bound on the  optimal value of $\mathbf {\bar{P}^{o}_2}$. 
Therefore, using the SCA technique, we propose an iterative algorithm to tighten the upper bound obtained from solving $\mathbf {\hat{P}^{o}_{2}}$, which is
summarized in Algorithm \ref{Alg:Alg2}.
Finally, we note that solving $\mathbf {\hat{P}^{o}_{2}}$  requires a complexity of
$\mathcal{O} \big( (2IJS+S)^3(3IJS+2IJ+2I+2S+J+1) \big)$ since there are $(2IJS+S)$ decision variables and $ (3IJS+2IJ+2I+2S+J+1)$ convex constraints in $\mathbf {\hat{P}^{o}_{2}}$~\cite{2020_TC_ClusterBasedResourceAllocationNPProbWith,2014_TWC_OptimaztionFramework3}.

\begin{algorithm}[t]
\caption{Iterative Approach for Resource Allocation with ASB in a PACSR}
\begin{algorithmic}[1] \label{Alg:Alg2}
\STATE \textbf{Initialization}: Set iteration count $\kappa=0$. Set initial point for $x^{(\kappa)}_{i,j,s}=0.5, \, \forall  i\in \mathcal {I}, j\in \mathcal {J}, s\in \mathcal {S}$.  Select a reasonably high penalty factor $\Lambda_2$ and low tolerance value $\epsilon$.
\WHILE{$ \sum\limits_{i\in \mathcal {I}} \sum\limits_{j\in \mathcal {J}} \sum\limits_{s\in \mathcal {S}} \Big (x_{i,j,s} \big (1 {-} 2 x^{(\kappa)}_{i,j,s} \big) {+} \big (x^{(\kappa)}_{i,j,s} \big)^{2} \Big)  \geqslant \epsilon$}
   \STATE Solve  \eqref{OptProb2Sol} using point $x^{(\kappa)}_{i,j,s}, \, \textrm{where}~ i\in \mathcal {I}, j\in \mathcal {J}, s\in \mathcal {S}$ and obtain solution parameters $P_{i,j,s}^{*}, x_{i,j,s}^{*} ,Z_{s}^*, \forall i\in \mathcal {I}, j\in \mathcal {J} , s\in \mathcal {S}$.
   \STATE Update point $x^{(\kappa+1)}_{i,j,s} = x^{*}_{i,j,s}, \forall i\in \mathcal {I}, j\in \mathcal {J}, s\in \mathcal {S}$.
   \STATE Update iteration count $\kappa = \kappa + 1$.
\ENDWHILE
\end{algorithmic} 
\end{algorithm}

\section{Numerical Results}
\label{Sec:Num}

\begin{table*}[t]
\caption{Value of System Parameters Used in Section~\ref{Sec:Num}.}\vspace{-11mm}
\begin{center}
\begin{tabular}{|l|l|l||l|l|l|}
\hline
\textsf{Parameter} & \textsf{Symbol}\!\!\!& \textsf{Value}  & \textsf{Parameter} & \textsf{Symbol}& \textsf{Value} \\
\hline
No. of users and MC order & $I$, $N$  &  6, 2   & No. of users supported by each AP& $M$& 3  \\ \hline
Heights of APs and users  & $h_{\textrm{A}}$, $h_{\textrm{U}}$  & $3.0~\textrm{m}$, $1.3~\textrm{m}$ & Upper bound on sub-band bandwidth & $B_{\textrm{max}}$ & $4.5~\textrm{GHz}$  \\
\hline
Height and radius of blockers \!\!\!\!\!& $h_{\textrm{B}}$, $r_{\textrm{B}}$ &  $1.7~\textrm{m}$,  $0.3~\textrm{m}$ & Gaurd band bandwidth  &  $B_{\textrm{g}}$  & $0.75~\textrm{GHz}$ \\ \hline
Density of wall blockers & $\lambda_{\textrm{B}}$   &   $0.2~\textrm{m}^{-2}$   &  Ratio of pulse duration to frame duration\!\!& $\varphi$ &  $1/2$ \\ \hline
Antenna gains &  $G_{\textrm{A}}$, $G_{\textrm{U}}\!\!\!$ &   $\!\!25~\textrm{dBi}$, $15~\textrm{dBi}\!\!\!$  & Path gain and rate thresholds per link & $L_{\textrm{thr}}$, $R_{\textrm{thr}}\!\!\!$ & $\!\!10^{-13}\!$, $2~\textrm{Gb/s}\!\!$  \\ \hline
 Noise spectral density & $N_{0}$ & $\!\!-174~\textrm{dBm/Hz}\!\!$     & Power budget per user& $P_{i}^{\textrm{max}}$ & 3.2~$\textrm{dBm}$         \\ \hline
\end{tabular}\label{tab1}
\end{center}\vspace{-7mm}
\end{table*}

In this section, we present numerical results to illustrate the performance of the two proposed resource allocation strategies. For clarity, we denote \emph{the proposed resource allocation with ESB} by \textsf{PRA1} and \emph{the proposed resource allocation with ASB in a PACSR} by  \textsf{PRA2}.  The numerical results are obtained by considering a rectangular indoor environment of size $20~\textrm{m}\times 20~\textrm{m}$. We consider that four APs are deployed symmetrically in the indoor environment in the same way as specified in the 3GPP standard~\cite{3GPPStand}.
Also, we consider that the 50 GHz spectrum that exists between $1.025$ and $1.075~\textrm{THz}$ in the first THz TW above 1 THz is used to serve the users\footnote{We clarify that it is indeed possible to utilize \textsf{PRA1} when the spectrum that exists anywhere within a THz TW is to be allocated. Despite this,  since \textsf{PRA2} can only be employed when the spectrum of interest  that is to be allocated exists in a PACSR/NACSR of the THz TW,  we present numerical results for \textsf{PRA1} and \textsf{PRA2} in a PACSR for fair comparison.}.
We use the absorption coefficient values that are calculated for the standard atmosphere with $10\%$ humidity~\cite{2011_Jornet_TWC}.
The values of the rest of the parameters used for numerical results are summarized in Table \ref{tab1}, unless specified otherwise\textcolor{black}{\footnote{\textcolor{black}{It is noted we consider that the sub-bands exist  only within TWs, where the variation of molecular absorption loss, as well as pulse broadening, is very small. Differently, a few previous studies, e.g., \cite{HBM3,2020_Chong_TWC_DistanceAdaptiveAbsorptionPeakModulation}, considered that the sub-bands can also exist within ACPRs, where the variation of molecular absorption loss, as well as pulse broadening, is very high. Due to this,  to avoid ISI, it is reasonable to consider $\varphi=1/2$ in this work, when \cite{HBM3,2020_Chong_TWC_DistanceAdaptiveAbsorptionPeakModulation} considered $\varphi=1/5$. }}}.

For \textsf{PRA2}, the values of the model parameters that are utilized in \eqref{Equ:CurveFit} to approximate the absorption coefficient values in the spectrum of interest are $\sigma_1=-90.996$, $\sigma_2=8.326\times 10^{-11}$, and $\sigma_3=0.0452$.
Moreover, in \textsf{PRA2},   for the substitution introduced in \eqref{Equ:Trans1}, we consider $\xi_{s}=5\times 10^{9},~\omega_{s}=0.5\times 10^9$, and $\varsigma_{s}=0.001$ $\forall$ $s\in \mathcal{S}$.
Furthermore, we note that Algorithm 1 and Algorithm 2, that correspond to \textsf{PRA1} and \textsf{PRA2}, respectively,  are implemented in AMPL 
 which is popular for modeling optimization problems\footnote{To cross-check, we implemented Algorithm 1 in Matlab CVX, and verified that the results obtained from AMPL well matches that obtained from Matlab CVX.}~\cite{AMPL1,Sheeraz2020}.
\textcolor{black}{Additionally, following \cite{Sheeraz2020,BinaryTrans2}, we consider  $\Lambda_1=200$, $\Lambda_2=200$, and $\epsilon= 10^{-6}$  in Algorithm 1 and Algorithm 2.}

We consider the state-of-the-art DAMC-based spectrum allocation  as the benchmark for \textsf{PRA1} and denote this strategy by \textsf{BM}~\cite{HBM2}.
Notably, as mentioned in Section I, \textsf{BM} uses ESB and considers that the  links with longer distances are assigned to center sub-bands and the  links with shorter distances are assigned to edge sub-bands~\cite{HBM2}.
Also, we consider \textsf{PRA1} as the benchmark for \textsf{PRA2} and note that while  \textsf{PRA1} considers spectrum allocation with ESB, \textsf{PRA2} considers spectrum allocation with ASB.

\begin{figure}[t]
\centering
\includegraphics[width=0.56\columnwidth]{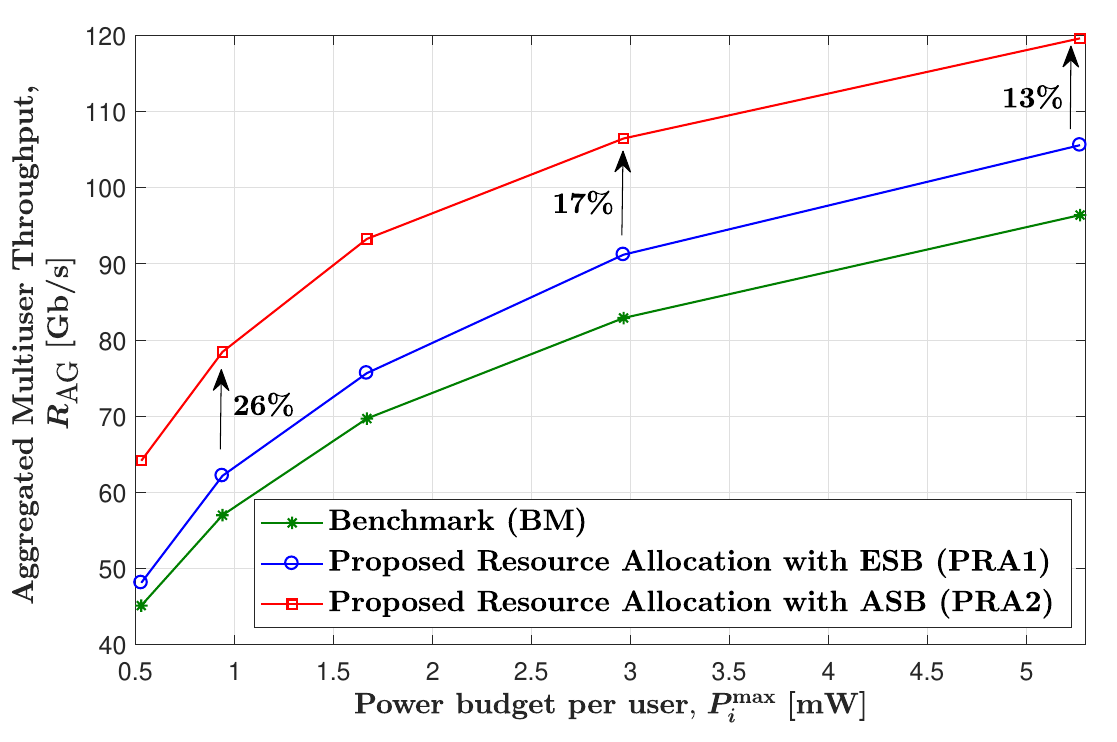}
\vspace{-3mm}
\caption{\textcolor{black}{Aggregated multiuser throughput versus the power budget per user.}}\label{NumFig:FigA}
\vspace{-2mm}
\end{figure}

In order to  demonstrate the performance improvement brought by  the proposed resource allocation strategies, we plot the aggregated multiuser throughput, $R_{\textrm{AG}}=\sum_{i=1}^{N}R_{i}$, versus the power budget per users for \textsf{BM}, \textsf{PRA1},   and \textsf{PRA2} in Fig. \ref{NumFig:FigA}.
We first observe that  $R_{\textrm{AG}}$ of \textsf{PRA1} is higher than that of \textsf{BM} for different user power budget levels.
This observation shows that the throughput fairness achieved by \textsf{BM} among associated links   does not necessarily guarantee the best throughput fairness among users in a MC-enabled multiuser THzCom system. It also shows that  a superior throughput fairness among the users can be achieved by utilizing  \textsf{PRA1}.
Next, as expected, we observe that \textsf{PRA2} achieves a significantly higher $R_{\textrm{AG}}$ compared to \textsf{PRA1} (between $13~\%$ and $26~\%$), due to the enhanced capabilities of \textsf{PRA2} relative to  \textsf{PRA1}. Specifically, while \textsf{PRA1} achieves a certain $R_{\textrm{AG}}$ with ESB,  the ASB capability of \textsf{PRA2} helps to further improve $R_{\textrm{AG}}$ by varying sub-band bandwidths.
These two observations demonstrate the significance of the optimal sub-band assignment  in MC-enabled multiuser THzCom systems and spectrum allocation with ASB, thereby demonstrating the benefits of our proposed resource allocation strategies.
We further observe that  $R_{\textrm{AG}}$ improves for \textsf{BM}, \textsf{PRA1}, and \textsf{PRA2} when the power budget increases. Interestingly, in the lower power budget regime where  all strategies achieve low $R_{\textrm{AG}}$, the $R_{\textrm{AG}}$ gain of \textsf{PRA2} relative to \textsf{PRA1} is higher, which shows that it is more beneficial to adopt  ASB 
when the power budget constraint becomes more stringent.

\begin{figure}[t]
\centering
   \includegraphics[width=0.56\columnwidth]{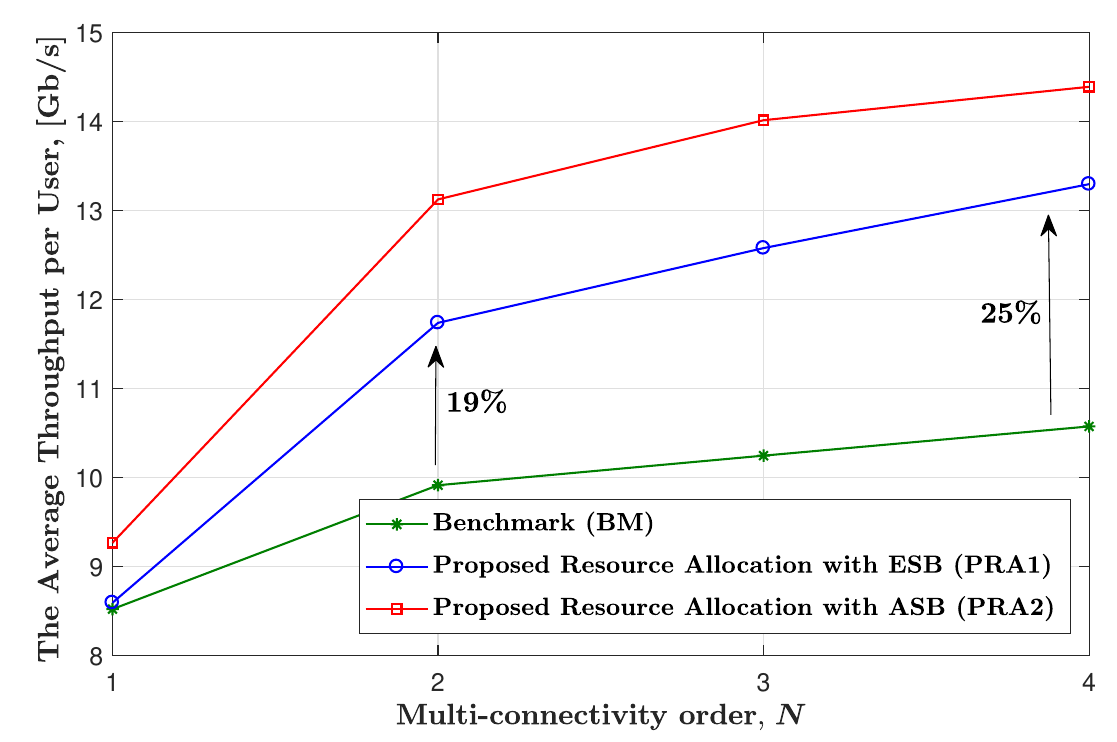}    \vspace{-4mm}
    \caption{\textcolor{black}{The average throughput per user versus the multi-connectivity order.}}\label{NumFig:FigD}
\end{figure}

In order to examine the impact of MC order on the proposed resource allocation strategies, we plot the average throughput per user versus the MC order for \textsf{BM}, \textsf{PRA1}, and \textsf{PRA2} in Fig. \ref{NumFig:FigD}.
We recall that the MC order is defined as the number of APs with which the users associate and communicate simultaneously.
For the sake of fairness, we keep the total number of sub-bands within the spectrum of interest unchanged. This is achieved   by changing the number of users in the system when the MC order is changed.
More precisely, the number of users corresponding to the MC orders of 1, 2, 3, and 4 are 12, 6, 4, and 3, respectively.
We first observe that the average throughput per user increases as MC order increases. This is because, although the power budget per user remains the same as MC order increases, the available bandwidth per user increases as MC order increases, thereby improving the throughput.
Second, we observe that although there exists a throughput gain for \textsf{PRA1} relative to \textsf{BM} for the MC orders of 2, 3, and 4, the gain greatly reduces when the MC order is 1.
This is due to the fact that the throughput fairness achieved by \textsf{BM} among associated links and the throughput fairness achieved by \textsf{PRA1} among users are the same when each user associates with one AP only, i.e., MC order is 1.
Third, we observe that the throughput gain for \textsf{PRA1} relative to \textsf{BM} increases as the MC order increases from 2 to 4. This is because, as MC order increases, \textsf{PRA1} gives more room to adapt the throughput among its associated links to achieve a higher throughput per user. However, as \textsf{BM} achieves throughput fairness among users by guaranteeing throughput fairness among links, the ability for a user to adapt throughput among its associated links does not exist when \textsf{BM} is employed. This leads to a low throughput improvement for \textsf{BM} as MC order increases, leading to a throughput gain for \textsf{PRA1} relative to \textsf{BM}. The second and third observations jointly show that it is more beneficial to employ \textsf{PRA1} when the MC order is high. Finally, we observe a steady throughput gain for \textsf{PRA2} relative to \textsf{PRA1}, which again shows the importance of spectrum allocation with ASB.

\begin{figure}[t]
\centering
\includegraphics[width=0.6\columnwidth]{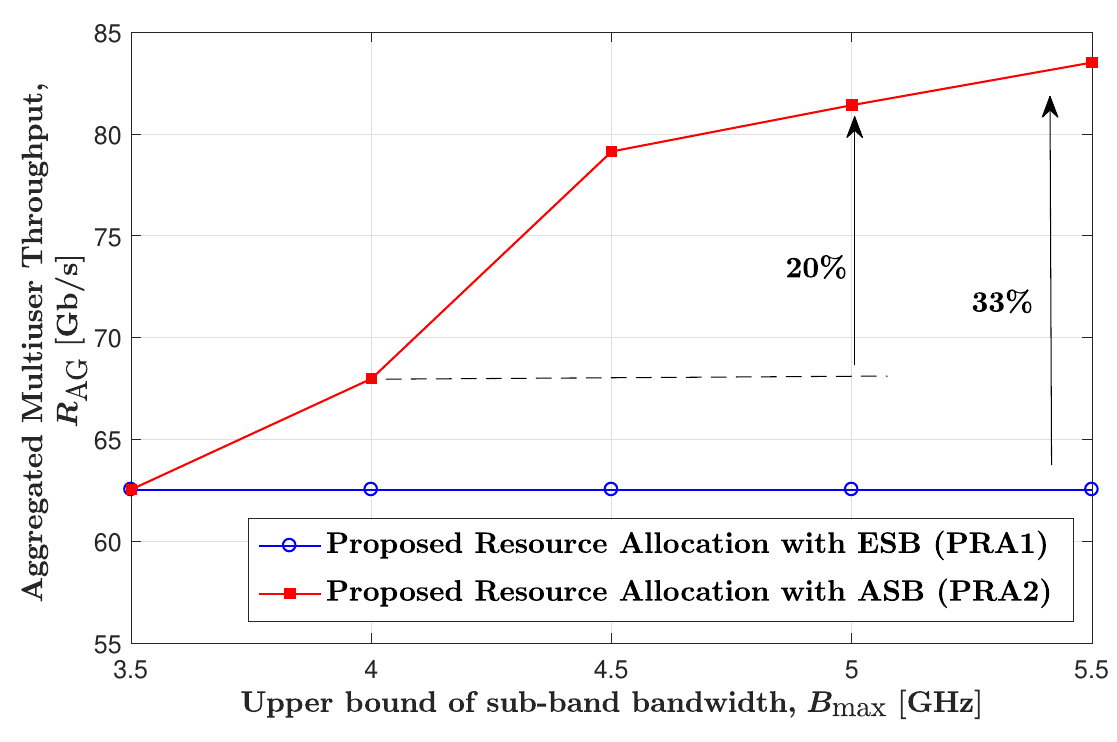}
   \vspace{-6mm}
\caption{Aggregated multiuser throughput versus the upper bound on the sub-band bandwidth.}\label{NumFig:FigB}
\end{figure}

In order to demonstrate the impact of the upper bound on the sub-band bandwidth, $B_{\textrm{max}}$, which is an important constraint in \textsf{PRA2}, we plot $R_{\textrm{AG}}$ versus $B_{\textrm{max}}$ for \textsf{PRA1} and \textsf{PRA2} in Fig. \ref{NumFig:FigB}.
We first observe that $R_{\textrm{AG}}$ of \textsf{PRA2} converges to that of \textsf{PRA1} when $B_{\textrm{max}}$ in \textsf{PRA2} is equal to the fixed sub-band bandwidth adopted in \textsf{PRA1}, such as $B_{\textrm{max}}=3.48~\textrm{GHz}$ in Fig. \ref{NumFig:FigB}. This shows the correctness of our proposed
resource allocation strategies.
We next observe a significant improvement in $R_{\textrm{AG}}$ when $B_{\textrm{max}}$ increases, e.g., a 20 $\%$ improvement in $R_{\textrm{AG}}$  when $B_{\textrm{max}}$ increases from $4~\textrm{GHz}$ to $5~\textrm{GHz}$. This is expected since a larger $B_{\textrm{max}}$ gives more room to exploit the ASB capability of \textsf{PRA2}, which improves the  $R_{\textrm{AG}}$ of \textsf{PRA2}.
Here, it should be noted that the  THz band amplifiers and transceivers capable of handling large bandwidth, which are required when $B_{\textrm{max}}$ is higher,  are in their infancy.
Also, pulses occupying  large bandwidths can be affected by increased pulse broadening  and channel squint effects due to the frequency selectivity of molecular absorption loss within sub-bands.
Thus, $B_{\textrm{max}}$ of THzCom systems must be carefully selected to achieve a trade-off between improving $R_{\textrm{AG}}$ performance and minimizing the complexity of software and hardware implementation.

\begin{figure}[t]
\centering
\includegraphics[width=0.6\columnwidth]{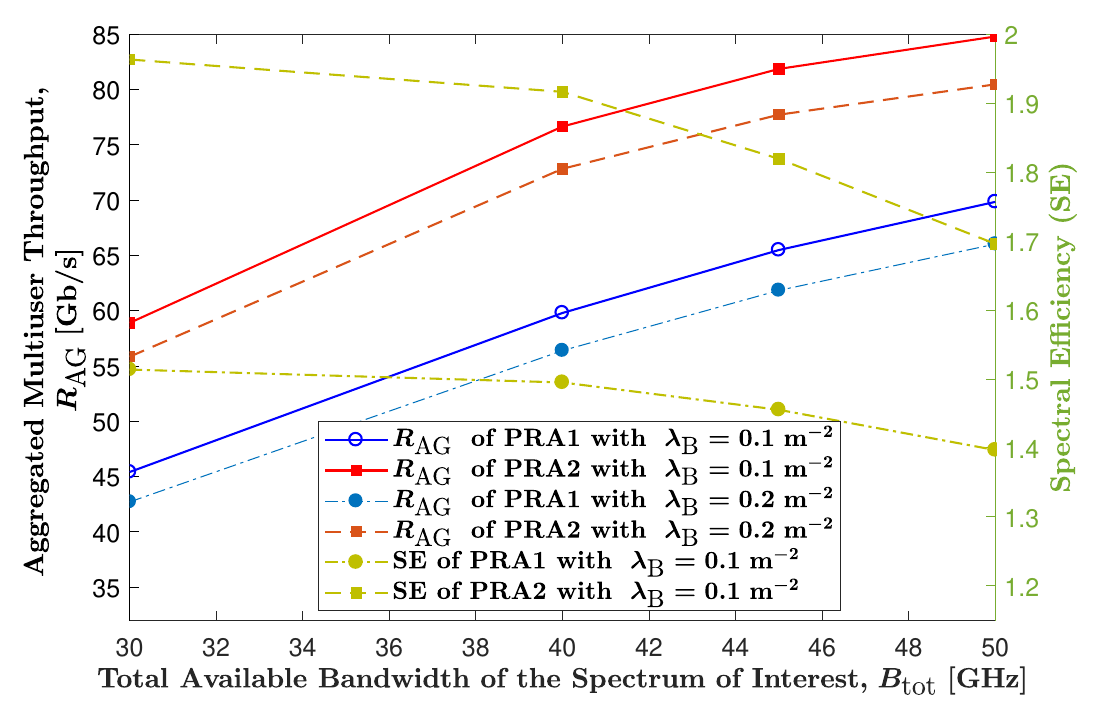}
\vspace{-6mm}
\caption{Aggregated multiuser throughput and spectral efficiency versus the total available bandwidth of the spectrum of interest for different blockage densities.}\label{NumFig:FigF}
\end{figure}

Fig. \ref{NumFig:FigF} plots $R_{\textrm{AG}}$ versus total bandwidth of the spectrum of interest, $B_{\textrm{tot}}$, for the proposed
resource allocation strategies at different densities of human blockers, $\lambda_{\textrm{B}}$.  Also, the spectral efficiency is plotted.
We first observe that the spectral efficiency of the proposed
resource allocation strategies decreases when $B_{\textrm{tot}}$ increases, due to the decrease in the power density per Hz when $B_{\textrm{tot}}$ increases.
We next observe that $R_{\textrm{AG}}$ of the proposed
resource allocation strategies increases when $B_{\textrm{tot}}$ increases.
This shows that the impact of increased availability of bandwidth for each user overwhelms the impact of decreasing spectral efficiency when $B_{\textrm{tot}}$ increases.
Moreover, we observe that  $R_{\textrm{AG}}$ of the proposed
resource allocation strategies decreases with increasing $\lambda_{\textrm{B}}$, due to the decrease in the time during which the links are available for data transmission when $\lambda_{\textrm{B}}$ increases.
This observation emphasizes the importance of carefully selecting the system parameters, e.g., power budget, MC order, maximum sub-band bandwidth, and total available bandwidth, for achieving the desired reliability and throughput performance, depending on the density of human blockers in the indoor environment.

\begin{figure}[t]
\centering
\includegraphics[width=0.58\columnwidth]{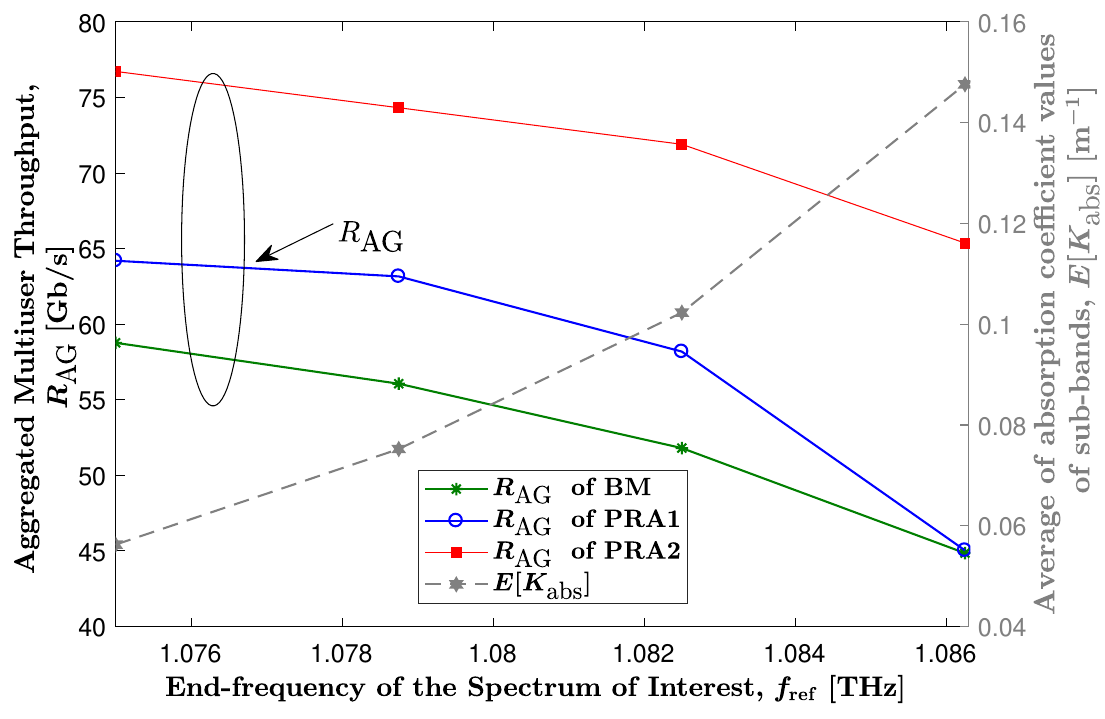}
\vspace{-6mm}
\caption{Aggregated multiuser throughput versus the end-frequency of the spectrum  of interest.}\label{NumFig:FigC}
\vspace{-2mm}
\end{figure}

We finally  investigate the impact of molecular absorption loss on the proposed resource allocation strategies. To  this end, we consider several 50 GHz spectra within the first THz TW above 1 THz, and plot  in Fig. \ref{NumFig:FigC} the $R_{\textrm{AG}}$ for \textsf{BM}, \textsf{PRA1}, and \textsf{PRA2}  within these spectra versus the end-frequency of these spectra, $f_{\textrm{ref}}$. 
Also, the average of the absorption coefficient values of the sub-bands within these spectra, $ E[K_{\textrm{abs}}]$, is plotted.
We first observe that $R_{\textrm{AG}}$ for \textsf{BM}, \textsf{PRA1}, and \textsf{PRA2} decreases when $f_{\textrm{ref}}$ increases. This is due to the increase in $ E[K_{\textrm{abs}}]$  when $f_{\textrm{ref}}$ increases, as shown in the figure, which in turn increases the molecular absorption loss in the sub-bands and reduces the reduction in $R_{\textrm{AG}}$.
We next observe  that the gain in $R_{\textrm{AG}}$ of \textsf{PRA1} relative to \textsf{BM} decreases when $f_{\textrm{ref}}$ increases. This observation can be explained as follows. As $f_{\textrm{ref}}$ increases, the number of sub-bands which have high molecular absorption loss increases. This leads to that the sub-band assignments where edge sub-bands allocated to links with shorter distances are the only sub-band assignments that can guarantee satisfying the constraint functions \eqref{OptProbOrg1-C} and \eqref{OptProbOrg-D} in $\mathbf {P^{o}_2}$. This in turn leads to the convergence of $R_{\textrm{AG}}$ of \textsf{PRA1} to that of \textsf{BM}.
These two observations indicate that in order to ensure improved performance, it is beneficial to select the spectrum within the THz TW that has the lowest average molecular absorption loss during spectrum allocation.

\section{Conclusions}

In this work, we investigated the impacts of  ASB and sub-band assignment in multi-band-based spectrum allocation on MC-enabled multiuser THzCom systems.  To this end, we formulated optimization problems with the primary focus on spectrum allocation.
Thereafter, we proposed reasonable approximations and transformations, and  developed iterative algorithms based on the SCA technique  to solve the formulated problems  analytically. Aided by numerical results, we showed that  by enabling ASB during spectrum allocation, the significantly higher throughput can be achieved as compared to adopting ESB and that this gain is most profound when the  power budget constraint is more stringent.  We also showed that the proposed sub-band assignment strategy in MC-enabled multiuser THzCom systems outperforms the  state-of-the-art sub-band assignment strategy and the performance gain is most profound when the spectrum with the lowest average molecular absorption loss within the THz TW is selected during spectrum allocation.

We clarify that the solution approach proposed in this work for the spectrum allocation with ASB is only applicable when the spectrum of interest  fully exists in either a PACSR or an NACSR of the THz band. Nevertheless, to  harness the potentials of the huge available bandwidths of THz TWs, it would be more beneficial to develop a solution for the spectrum allocation with ASB when the spectrum of interest  exists anywhere within a THz TW, which will be considered in our future work.

\appendices

\section{Proof of Lemma 1}\label{app:Derive_Lemma1}

We first prove the concavity of  $|\alpha_{i,j,s}|^2$ w.r.t. to $Z_{\nu}$, where $i\in \mathcal{I},j\in \mathcal{J}, s\in \mathcal{S},\nu\in \mathcal{S}$.
To this end, we rearrange $|\alpha_{i,j,s}|^2$ given in \eqref{Equ:AG21} to obtain
\begin{align} \label{Equ:Lem1}
 |\alpha_{s}|^2&= \varrho  ~\! \frac{ e^{-\hat{K}(f_s) d}}{\left(f_s d \right)^2} =\varrho_1 ~\!\frac{e^{- \Phi_{s,{\nu}}e^{-b_{s,{\nu}}\log(\varsigma_{\nu} Z_{{\nu}})} }}{(\Omega_{s,{\nu}}-a_{s,{\nu}}\omega_{\nu} \log(\varsigma_{\nu} Z_{{\nu}}))^{2}}, ~~\forall s\in \mathcal{S},
\end{align}

\noindent where $\varrho_1=\varrho e^{-d\sigma_3}/d^2 $, $\textcolor{black}{\Phi_{s,{\nu}}}=d(e^{\sigma_{1} +\sigma_{2}\Omega_{s,{\nu}}})$, $\Omega_{s,{\nu}}=\digamma_{s}- \sum\limits_{k\in \mathcal {S}}a_{s,k}\xi_{k}-\sum\limits_{k\in \mathcal {S}/{\nu}}\omega_{k}\log(\varsigma_{k} Z_{k})$, and $b_{s,{\nu}}=\sigma_{2}a_{s,{\nu}}\omega_{\nu}$. We note that the subscript $i$ and $j$ are dropped in \eqref{Equ:Lem1} for brevity.
 Thereafter, we take the second derivative of \eqref{Equ:Lem1} w.r.t. to $Z_{\nu}$ and rearrange the resulting terms to arrive at
\begin{align}
  \frac{\partial^2(|\alpha_{s}|^2)}{\partial Z^2_{\nu}}
 &= \begin{cases}
E^{s,\nu}_{1,1} , & {\nu} \leqslant  s, \\
E^{s,\nu}_{1,2}, & \textrm{elsewhere},\end{cases} ~s\in \mathcal{S},\nu\in \mathcal{S},
\end{align}
where $E^{s,\nu}_{1,1}=- \varrho_1|\alpha_{s}|^2 b_{s,{\nu}}^2\left(\left(1+1/b_{s,{\nu}}\right)-d \hat{K}(f_s)e^{d\sigma_3}\right)d \hat{K}(f_s)e^{d\sigma_3}\Omega_{s,{\nu}}^{-2}Z_{\nu }^{-2}$ and $E^{s,\nu}_{1,2}=0$.
It can be shown that $E^{s,\nu}_{1,1} \leqslant 0$, $\forall$ $s\in \mathcal{S},\nu\in \mathcal{S}$ when $1/\omega_{\nu}>\bar{\omega}$, $\forall$ $\nu\in \mathcal{S}$. Based on this, we conclude that $|\alpha_{i,j,s}|^2$ in \eqref{Equ:AG21} is concave w.r.t. to $Z_{\nu}$ when $1/\omega_{\nu}>\bar{\omega}$.

We next prove the concavity of $R_{i,j,s}^{\textrm{nb}}$ w.r.t. to $Z_{\nu}$, where $i\in \mathcal{I},j\in \mathcal{J}, s\in \mathcal{S},\nu\in \mathcal{S}$. To this end, we rearrange $R_{i,j,s}^{\textrm{nb}}$ given in \eqref{Equ:Rateijs22} to obtain $ R_{s}^{\textrm{nb}}= B_s(Z_s) \bar{\varphi} \log \left(1+\Upsilon_s^{\nu}\right)$,
where $B_s(Z_s)=\xi_s+\omega_s\log(\varsigma_s Z_{s})$, $\bar{\varphi}=\frac{\varphi}{\log(2)}$, and $\textcolor{black}{\Upsilon_s^{\nu}}=\frac{P_{s}G_{\textrm{A}} G_{\textrm{U}}}{ N_{0}\Omega^2_{s,{\nu}}B_s(Z_s)}~\!e^{- \Phi_{s,{\nu}}e^{-b_{s,{\nu}}\log(\varsigma_{\nu} Z_{{\nu}})} }$.
 Thereafter, we take the second derivative of $R_{s}^{\textrm{nb}}$ w.r.t. to $Z_{\nu}$ and rearrange the resulting terms to arrive at
 \begin{align} 
  &\frac{\partial^2R_{s}^{\textrm{nb}}}{\partial Z^2_{\nu}}= \begin{cases}
E^{s,\nu}_{2,1}, & {\nu} <  s, \\
E^{s,\nu}_{2,2}& {\nu} =  s, \\
E^{s,\nu}_{2,3}, & \textrm{elsewhere},\end{cases} ~\forall s\in \mathcal{S},\nu\in \mathcal{S},
\end{align}
where
\begin{align} 
E^{s,\nu}_{2,1}&=-\dfrac{B_s(Z_s)b_{s,{\nu}}^2\Upsilon_s^{\nu}\bar{\varphi}\left(\left(1+\Upsilon_s^{\nu}\right)\left(1+1/b_{s,{\nu}}\right) -d~\hat{K}(f_s)e^{d\sigma_3}\right)d~\hat{K}(f_s)e^{d\sigma_3}}{Z_{\nu}^2\left(1+\Upsilon_s^{\nu}\right)^2},
\end{align}
\begin{align} \label{Eq:Esv2}
E^{s,\nu}_{2,2}&=\underbrace{-\dfrac{{\bar{\varphi}\textcolor{black}{\omega_{s}}}}{Z_{s}^2}\left(\log\left(1+\Upsilon_s^{s}\right)-\dfrac{\Upsilon_s^{s}}{\left(1+\Upsilon_s^{s}\right)}\right)}_{\substack{T_{1}}}\underbrace{-\dfrac{(\Upsilon_s^{s})^2\bar{\varphi}\left(b_{s,{s}}B_s(Z_s)d~\hat{K}(f_s)e^{d\sigma_3}-{\textcolor{black}{\omega_s}}\right)^2}{\left(1+\Upsilon_s^{s}\right)^2Z_{s}^2B_s(Z_s)} }_{\substack{T_{2}}} \notag \\
&~~~~~~~~\underbrace{-\dfrac{b^2_{s,{s}}\Upsilon_s^{s}B_s(Z_s)\bar{\varphi}\left((1+1/b_{s,{s}})-d~\hat{K}(f_s)e^{d\sigma_3}\right)d~\hat{K}(f_s)e^{d\sigma_3}}{Z_{s}^2(1+\Upsilon_s^{s})}}_{\substack{T_{3}}},
\end{align}
and $E^{s,\nu}_{2,3}=0$. 
It can be shown that $E^{s,\nu}_{2,1}<0$ when $1/\omega_{\nu}>\bar{\omega}$. Also, it can be shown that $E^{s,\nu}_{2,2}<0$ when $1/\omega_{\nu}>\bar{\omega}$, since $T_1<0$ 
and $T_2<0$, $\forall$ $s \in \mathcal{S},\nu\in \mathcal{S}$ and $T_3<0$ when $1/\omega_{\nu}>\bar{\omega}$. 
Based on these, we conclude that $R_{i,j,s}^{\textrm{nb}}$ in \eqref{Equ:Rateijs22} is concave w.r.t. to $Z_{\nu}$ when $1/\omega_{\nu}>\bar{\omega}$.

 \bibliographystyle{IEEEtran}
 \bibliography{TCOM2021R1V2ArXiv}

\begin{thebibliography}{10}
\providecommand{\url}[1]{#1}
\csname url@samestyle\endcsname
\providecommand{\newblock}{\relax}
\providecommand{\bibinfo}[2]{#2}
\providecommand{\BIBentrySTDinterwordspacing}{\spaceskip=0pt\relax}
\providecommand{\BIBentryALTinterwordstretchfactor}{4}
\providecommand{\BIBentryALTinterwordspacing}{\spaceskip=\fontdimen2\font plus
\BIBentryALTinterwordstretchfactor\fontdimen3\font minus
  \fontdimen4\font\relax}
\providecommand{\BIBforeignlanguage}[2]{{%
\expandafter\ifx\csname l@#1\endcsname\relax
\typeout{** WARNING: IEEEtran.bst: No hyphenation pattern has been}%
\typeout{** loaded for the language `#1'. Using the pattern for}%
\typeout{** the default language instead.}%
\else
\language=\csname l@#1\endcsname
\fi
#2}}
\providecommand{\BIBdecl}{\relax}
\BIBdecl

\bibitem{2020_Mag6G_Marco_UseCasesandTechnologies}
M.~{Giordani}, M.~{Polese}, M.~{Mezzavilla}, S.~{Rangan}, and M.~{Zorzi},
  ``Toward {6G} networks: Use cases and technologies,'' \emph{IEEE Commun.
  Mag.}, vol.~58, no.~3, pp. 55--61, Mar. 2020.

\bibitem{2018MagCombatDist}
I.~F. {Akyildiz}, C.~{Han}, and S.~{Nie}, ``Combating the distance problem in
  the millimeter wave and terahertz frequency bands,'' \emph{IEEE Commun.
  Mag.}, vol.~56, no.~6, pp. 102--108, June 2018.

\bibitem{2020_WCM_THzMag_TerahertzNetworks}
M.~{Polese}, J.~M. {Jornet}, T.~{Melodia}, and M.~{Zorzi}, ``Toward end-to-end,
  full-stack {6G} terahertz networks,'' \emph{IEEE Commun. Mag.}, vol.~58,
  no.~11, pp. 48--54, Nov. 2020.

\bibitem{2020_WCM_THzMag_Standardization}
V.~{Petrov}, T.~{Kurner}, and I.~{Hosako}, ``{IEEE} 802.15.3d: First
  standardization efforts for sub-terahertz band communications toward 6{G},''
  \emph{IEEE Commun. Mag.}, vol.~58, no.~11, pp. 28--33, Nov. 2020.

\bibitem{2011_Jornet_TWC}
J.~M. {Jornet} and I.~F. {Akyildiz}, ``Channel modeling and capacity analysis
  for electromagnetic wireless nanonetworks in the terahertz band,'' \emph{IEEE
  Trans. Wireless Commun.}, vol.~10, no.~10, pp. 3211--3221, Oct. 2011.

\bibitem{Chong2019Archive}
\BIBentryALTinterwordspacing
C.~Han, Y.~Wu, Z.~Chen, and X.~Wang, ``Terahertz communications ({TeraCom}):
  Challenges and impact on {6G} wireless systems,'' Dec. 2019. [Online].
  Available: \url{https://arxiv.org/abs/1912.06040}
\BIBentrySTDinterwordspacing

\bibitem{akramICC2020}
A.~Shafie, N.~Yang, and C.~Han, ``Multi-connectivity for indoor terahertz
  communication with self and dynamic blockage,'' in \emph{Proc. IEEE Int.
  Conf. Commun. (ICC)}, Dublin, Ireland, June 2020, pp. 1--7.

\bibitem{HBM1}
Z.~{Hossain} and J.~M. {Jornet}, ``Hierarchical bandwidth modulation for
  ultra-broadband terahertz communications,'' in \emph{Proc. IEEE ICC},
  Shanghai, China, May 2019, pp. 1--7.

\bibitem{HBM3}
C.~{Han}, A.~O. {Bicen}, and I.~F. {Akyildiz}, ``Multi-wideband waveform design
  for distance-adaptive wireless communications in the terahertz band,''
  \emph{IEEE Trans. Signal Process.}, vol.~64, no.~4, pp. 910--922, Feb. 2016.

\bibitem{HBM2}
C.~{Han} and I.~F. {Akyildiz}, ``Distance-aware bandwidth-adaptive resource
  allocation for wireless systems in the terahertz band,'' \emph{IEEE Trans.
  THz Sci. Technol.}, vol.~6, no.~4, pp. 541--553, July 2016.

\bibitem{2019_Chong_DABM2}
X.~{Zhang}, C.~{Han}, and X.~{Wang}, ``Joint beamforming-power-bandwidth
  allocation in terahertz {NOMA} networks,'' in \emph{Proc. Int. Conf. Sensing,
  Commun., Netw. (SECON)}, Boston, MA, USA, Sept. 2019, pp. 1--9.

\bibitem{2020_Chong_InfoCom_DABM}
M.~{Yu}, A.~{Tang}, X.~{Wang}, and C.~{Han}, ``Joint scheduling and power
  allocation for {6G} terahertz mesh networks,'' in \emph{Proc. Int. Conf.
  Comput. Netw. Commun. (ICNC)}, Big Island, HI, USA, Feb. 2020, pp. 631--635.

\bibitem{2021_THz_NOMA}
H.~Zhang, Y.~Duan, K.~Long, and V.~C.~M. Leung, ``Energy efficient resource
  allocation in terahertz downlink {NOMA} systems,'' \emph{IEEE Trans.
  Commun.}, vol.~69, no.~2, pp. 1375--1384, Feb. 2021.

\bibitem{2020ICC_NOMAforTHz}
H.~Zhang, H.~Zhang, W.~liu, K.~long, J.~Dong, and V.~C.~M. Leung, ``Energy
  efficient user clustering and hybrid precoding for terahertz {MIMO}-{NOMA}
  systems,'' in \emph{Proc. IEEE Int. Conf. Commun. (ICC)}, Dublin, Ireland,
  June 2020, pp. 1--5.

\bibitem{2020_Chong_IRSTHz2}
X.~Ma, Z.~Chen, W.~Chen, Z.~Li, Y.~Chi, C.~Han, and S.~Li, ``Joint channel
  estimation and data rate maximization for intelligent reflecting surface
  assisted terahertz {MIMO} communication systems,'' \emph{IEEE Access},
  vol.~8, pp. 99\,565--99\,581, May 2020.

\bibitem{2021_THzMCHandOver}
M.~F. \"Ozkoc, A.~Koutsaftis, R.~Kumar, P.~Liu, and S.~S. Panwar, ``The impact
  of multi-connectivity and handover constraints on millimeter wave and
  terahertz cellular networks,'' \emph{IEEE J. Sel. Areas Commun.}, vol.~39,
  no.~6, pp. 1833--1853, June 2021.

\bibitem{2020_TC_MLApproach_GeoffreyYeLi}
R.~{Liu}, M.~{Lee}, G.~{Yu}, and G.~Y. {Li}, ``User association for
  millimeter-wave networks: A machine learning approach,'' \emph{IEEE Trans.
  Commun.}, vol.~68, no.~7, pp. 4164--4174, July 2020.

\bibitem{2020_IEEESurveyandTutorials_MCforURLLC}
M.~{Suer}, C.~{Thein}, H.~{Tchouankem}, and L.~{Wolf}, ``Multi-connectivity as
  an enabler for reliable low latency communications-an overview,'' \emph{IEEE
  Commun. Surveys Tuts.}, vol.~22, no.~1, pp. 156--169, Firstquarter 2020.

\bibitem{2018_TC_Idea}
Y.~{Liu}, X.~{Fang}, M.~{Xiao}, and S.~{Mumtaz}, ``Decentralized beam pair
  selection in multi-beam millimeter-wave networks,'' \emph{IEEE Trans.
  Commun.}, vol.~66, no.~6, pp. 2722--2737, June 2018.

\bibitem{2017N5}
A.~{Moldovan}, P.~{Karunakaran}, I.~F. {Akyildiz}, and W.~H. {Gerstacker},
  ``Coverage and achievable rate analysis for indoor terahertz wireless
  networks,'' in \emph{Proc. IEEE Int. Conf. Commun. (ICC)}, Paris, France, May
  2017, pp. 1--7.

\bibitem{2016_Jornet_MultiRAT}
X.-W. Yao and J.~M. Jornet, ``{TAB-MAC}: Assisted beamforming {MAC} protocol
  for terahertz communication networks,'' \emph{Nano Commun. Netw.}, vol.~9,
  pp. 36--42, Sept. 2016.

\bibitem{2020_WCL_Multi_RATMCforTHz}
J.~{Sayehvand} and H.~{Tabassum}, ``Interference and coverage analysis in
  coexisting {RF} and dense terahertz wireless networks,'' \emph{IEEE Wireless
  Commun. Lett.}, vol.~9, no.~10, pp. 1738--1742, Oct. 2020.

\bibitem{3GPPStand}
{3GPP}, ``Study on channel model for frequency spectrum above 6 {GHz},'' 3GPP
  TR 38.900 V14.2.0, June 2017.

\bibitem{akram2020JSAC}
A.~Shafie, N.~Yang, S.~Durrani, X.~Zhou, C.~Han, and M.~Juntti, ``Coverage
  analysis for 3{D} terahertz communication systems,'' \emph{IEEE J. Sel. Areas
  Commun.}, vol.~39, no.~6, pp. 1817--1832, June 2021.

\bibitem{akramICCWS2020}
A.~Shafie, N.~Yang, Z.~Sun, and S.~Durrani, ``Coverage analysis for {3D}
  terahertz communication systems with blockage and directional antennas,'' in
  \emph{Proc. IEEE Int. Conf. Commun. (ICC) Workshop}, Dublin, Ireland, June
  2020, pp. 1--7.

\bibitem{RDM1}
P.~Nain, D.~Towsley, B.~Liu, and Z.~Liu, ``Properties of random direction
  models,'' in \emph{Proc. IEEE Conf. Comput. Commun. Workshops (INFOCOM
  Wkshps)}, Miami, FL, Mar. 2005, pp. 1897--1907.

\bibitem{MC4}
M.~{Gapeyenko}, V.~{Petrov}, D.~{Moltchanov}, M.~R. {Akdeniz}, S.~{Andreev},
  N.~{Himayat}, and Y.~{Koucheryavy}, ``On the degree of multi-connectivity in
  5{G} millimeter-wave cellular urban deployments,'' \emph{IEEE Trans. Veh.
  Technol.}, vol.~68, no.~2, pp. 1973--1978, Feb. 2019.

\bibitem{2021_JSAC_VariableBandwidth}
A.~Saeed, O.~Gurbuz, A.~O. Bicen, and M.~A. Akkas, ``Variable-bandwidth model
  and capacity analysis for aerial communications in the terahertz band,''
  \emph{IEEE J. Sel. Areas Commun.}, vol.~39, no.~6, pp. 1768--1784, June 2021.

\bibitem{2011_TWC_IBISuppression}
M.~Ma, X.~Huang, B.~Jiao, and Y.~J. Guo, ``Optimal orthogonal precoding for
  power leakage suppression in {DFT}-based systems,'' \emph{IEEE Trans.
  Commun.}, vol.~59, no.~3, pp. 844--853, Mar. 2011.

\bibitem{2008_Hitran}
L.~Rothman, I.~Gordon, A.~Barbe, D.~Benner, P.~Bernath, M.~Birk, V.~Boudon, and
  L.~Brown, ``The {HITRAN} 2008 molecular spectroscopic database,'' \emph{J.
  Quantitative Spectroscopy Radiative Transfer}, vol. 110, no.~9, pp. 533--572,
  July 2009.

\bibitem{2017_ChongTVT_Graphene}
C.~{Han} and I.~F. {Akyildiz}, ``Three-dimensional end-to-end modeling and
  analysis for graphene-enabled terahertz band communications,'' \emph{IEEE
  Trans. Veh. Technol.}, vol.~66, no.~7, pp. 5626--5634, July 2017.

\bibitem{MC1}
V.~{Petrov}, D.~{Solomitckii}, A.~{Samuylov}, M.~A. {Lema}, M.~{Gapeyenko},
  D.~{Moltchanov}, S.~{Andreev}, V.~{Naumov}, K.~{Samouylov}, M.~{Dohler}, and
  Y.~{Koucheryavy}, ``Dynamic multi-connectivity performance in ultra-dense
  urban {mmWave} deployments,'' \emph{IEEE J. Sel. Areas Commun.}, vol.~35,
  no.~9, pp. 2038--2055, Sep. 2017.

\bibitem{2020_Chong_TWC_DistanceAdaptiveAbsorptionPeakModulation}
W.~{Gao}, Y.~{Chen}, C.~{Han}, and Z.~{Chen}, ``Distance-adaptive absorption
  peak modulation ({DA-APM}) for terahertz covert communications,'' \emph{IEEE
  Trans. Wireless Commun.}, vol.~20, no.~3, pp. 2064--2077, Nov. 2020.

\bibitem{HangNewChina}
H.~Yuan, X.~Wang, K.~Yang, and J.~An, ``Hybrid precoding for cluster-based
  multi-carrier beam division multiple access in terahertz wireless
  communications,'' \emph{China Commun.}, vol.~18, no.~5, pp. 81--92, May 2021.

\bibitem{2019_TC_JointUserAssociationandResourceAllocation_GeoffreyYeLi_subbandSC}
R.~{Liu}, Q.~{Chen}, G.~{Yu}, and G.~Y. {Li}, ``Joint user association and
  resource allocation for multi-band millimeter-wave heterogeneous networks,''
  \emph{IEEE Trans. Commun.}, vol.~67, no.~12, pp. 8502--8516, Dec. 2019.

\bibitem{ConvexBoyedBook}
S.~Boyd and L.~Vandenberghe, \emph{Convex Optimization}.\hskip 1em plus 0.5em
  minus 0.4em\relax Cambridge, U.K.: Cambridge Univ. Press, 2004.

\bibitem{2020_WCNC_HangNan}
H.~{Yuan}, N.~{Yang}, K.~{Yang}, C.~{Han}, and J.~{An}, ``Enabling massive
  connections using hybrid beamforming in terahertz micro-scale networks,'' in
  \emph{IEEE Wireless Commun. Netw. Conf. (WCNC)}, Seoul, South Korea, June
  2020, pp. 1--7.

\bibitem{2014_PenaltyFactor2}
E.~{Che}, H.~D. {Tuan}, and H.~H. {Nguyen}, ``Joint optimization of cooperative
  beamforming and relay assignment in multi-user wireless relay networks,''
  \emph{IEEE Trans. Wireless Commun.}, vol.~13, no.~10, pp. 5481--5495, Oct.
  2014.

\bibitem{BinaryTrans1}
D.~W.~K. {Ng} and R.~{Schober}, ``Secure and green {SWIPT} in distributed
  antenna networks with limited backhaul capacity,'' \emph{IEEE Trans. Wireless
  Commun.}, vol.~14, no.~9, pp. 5082--5097, Sept. 2015.

\bibitem{BinaryTrans2}
U.~{Rashid}, H.~D. {Tuan}, H.~H. {Kha}, and H.~H. {Nguyen}, ``Joint
  optimization of source precoding and relay beamforming in wireless {MIMO}
  relay networks,'' \emph{IEEE Trans. Commun.}, vol.~62, no.~2, pp. 488--499,
  Feb. 2014.

\bibitem{Sheeraz2020}
S.~A. {Alvi}, X.~{Zhou}, S.~{Durrani}, and D.~T. {Ngo}, ``Sequencing and
  scheduling for multi-user machine-type communication,'' \emph{IEEE Trans.
  Commun.}, vol.~68, no.~4, pp. 2459--2473, Apr. 2020.

\bibitem{CVX}
M.~Grant and S.~Boyd, ``{CVX}: Matlab software for disciplined convex
  programming, v2.1,'' http://cvxr.com/cvx, Mar. 2014.

\bibitem{2020_TC_ClusterBasedResourceAllocationNPProbWith}
B.~{Soleimani} and M.~{Sabbaghian}, ``Cluster-based resource allocation and
  user association in mmwave femtocell networks,'' \emph{IEEE Trans. Commun.},
  vol.~68, no.~3, pp. 1746--1759, Mar. 2020.

\bibitem{2014_TWC_OptimaztionFramework3}
E.~{Che}, H.~D. {Tuan}, and H.~H. {Nguyen}, ``Joint optimization of cooperative
  beamforming and relay assignment in multi-user wireless relay networks,''
  \emph{IEEE Trans. Wireless Commun.}, vol.~13, no.~10, pp. 5481--5495, Oct.
  2014.

\bibitem{2015_TC_BA2_forBminArguement}
Z.~{Wang}, V.~{Aggarwal}, and X.~{Wang}, ``Joint energy-bandwidth allocation in
  multiple broadcast channels with energy harvesting,'' \emph{IEEE Trans.
  Commun.}, vol.~63, no.~10, pp. 3842--3855, Oct. 2015.

\bibitem{AMPL1}
R.~Fourer, D.~Gay, and B.~Kernighan, ``{AMPL}: A mathematical programing
  language,'' in \emph{Algorithms and Model Formulations in Mathematical
  Programming. ({NATO} {ASI} Series, Series {F}: Computer and Systems
  Sciences)}, vol.~51, S. W. Wallace, Ed. Berlin, Germany: Springer, 1989.

\end{thebibliography}
\end{document}